\documentclass[paper,10pt]{JHEP3}

\usepackage{amsmath}
\usepackage{prelim2e}
\usepackage{feynmp}
\usepackage{graphicx}

\allowdisplaybreaks[1]

\author{
T.G. Birthwright, E.~W.~N.~Glover and P. Marquard  \\
Department of Physics,
University of Durham,
Durham DH1 3LE,
England\\
E-mail:  \email{T.G.Birthwright@durham.ac.uk, E.W.N.Glover@durham.ac.uk,
Peter.Marquard@durham.ac.uk}}

\title{Master Integrals For Massless Two-Loop Vertex Diagrams With 
Three Offshell Legs}

\preprint{DCPT/04/90, IPPP/04/45, hep-ph/0407343}

\abstract{We compute the master integrals for massless two-loop vertex graphs
with three off-shell legs.  These master integrals are relevant for the QCD
corrections to  $H \to V^*V^*$ (where $V = W$, $Z$)  and for two-loop studies
of the triple gluon (and quark-gluon) vertex. We employ the differential
equation technique to  provide series expansions in $\epsilon$ for the various
master integrals.  The results are analytic and  contain a new class of two-dimensional
harmonic polylogarithms. 
}

\begin{document}

\newcommand{\Mvariable}[1]{{\mathrm{ #1}}}
\newcommand{\Muserfunction}[1]{{\mathrm{ #1}}}

\newcommand{\FFF}{\ensuremath{{}_2F_1}}
\newcommand{\BB}{\mathrm{BB}}
\newcommand{\GL}{\mathrm{GL}}
\newcommand{\TGL}{\mathrm{TGL}}
\newcommand{\TB}{\mathrm{TB}}
\newcommand{\SSS}{\mathrm{SS}}

\begin{fmffile}{pics}

\section{Introduction}
\label{sec:intro}

It is well known that many of the loop integrals  that appear in Feynman
diagram calculations can be expressed in terms of hypergeometric functions 
with parameters that depend on the number of space time dimensions $d$ and a
number of kinematic scales. However, expressing these hypergeometric functions
as expansions in  $\epsilon = (4-d)/2$ is rather non-trivial.  In general the
coefficients  involve polylogarithms, both of the Nielsen~\cite{kolbig} and
harmonic~\cite{RV,GR} varieties, and often new polylogarithms need to be
introduced.  

Integration by parts~\cite{IBP1,IBP2} (IBP) and Lorentz
invariance~\cite{diffeqLI} (LI) identities are crucial in
reducing~\cite{Laporta} the number of master integrals (MI)  that actually need
to be evaluated and  several powerful tools have been established to deal with
the problem.   Often, these methods rely on the link between the hypergeometric
functions that yield (nested) sums and their integral representations that
yield polylogarithms. Two of the most powerful analytic methods are the
Mellin-Barnes technique~\cite{BD} and the differential equations
approach~\cite{Kotikov}. Both have been used extensively to provide expansions
in $\epsilon$ for  two-loop box graphs with massless internal propagators when
all the external legs are
on-shell~\cite{planarA,nonplanarA,planarB,nonplanarB,planarIR} and when one of
the  external legs is off-shell~\cite{GRplanar,GRnonplanar}. All of these
expansions have been checked by Binoth and Heinrich's numerical
program~\cite{BHnumer} for evaluating loop integrals.  Once analytic expansions
for the  two-loop master integrals were known, they were rapidly exploited in
the calculation of the amplitudes for  physical scattering processes.   

The two-loop helicity amplitudes have been evaluated for the 
gluon-gluon~\cite{BDKgggg,BFDgggg}, quark-gluon~\cite{BFDqqgg,qqgghel} and
quark-quark~\cite{qqqqhel} processes and have confirmed the earlier ``squared"
matrix elements~\cite{qqunlike,qqlike,qqgg,gggg}. Similarly, amplitudes  for
the phenomenologically important $gg \to \gamma\gamma$~\cite{BFDggpp}  and
$q\bar q \to \gamma\gamma$~\cite{qqpp}  processes as well as $\gamma\gamma \to
\gamma\gamma$~\cite{BFDGpppp,BGMBpppp} and (massless) Bhabha
scattering~\cite{BDGbha} have also been computed. The processes with one
off-shell leg include $e^+e^- \to 3$~jets~\cite{3jme,helamp3j,MUW3j} which is
crucial in making a precise  determination of the strong coupling constant at
the NLC.   Progress is also being made in calculating the two-loop QED
corrections to Bhabha scattering which is of crucial importance in determining
the luminosity  at the NLC (see the nice review of the current status in
Ref.~\cite{czakon}). Here there are 33 double box graphs to evaluate of which
seven have been studied~\cite{bhabha1,bhabha2,bhabha3,czakon}.   Analytic
expressions for the associated vertex graphs are also
known~\cite{bhabhavertex,czakon} and have been employed to calculate the
QED~\cite{qedff} and QCD~\cite{qcdff} corrections to the massive fermion form
factor. 

In this paper, we take a first step towards calculating massless two-loop 
$2\to 2$ scattering amplitudes with two off-shell legs. These  processes
include the NNLO QCD corrections to $q\bar q \to V^* V^*$ (where $V = W$,$Z$)
and the NLO corrections to $gg \to V^*V^*$.  Here we indicate that the gauge
bosons are off shell to account for resonance effects in the decay of the gauge
boson.  As in the on-shell and single off-shell cases, the MI include planar
and non-planar box graphs as well as vertex and self energies.   Altogether
there are 11 planar box and 3 non-planar box master topologies, some of which
require 2 or more MI to be computed.   Our goal in this paper is much more
modest.   As a first step, we restrict ourselves to the planar and non-planar
vertex graphs that form a simpler subset of the necessary MI.   

The MI presented here are ingredients for several interesting two-loop
processes in their own right.  In Higgs physics, the $H \to V^*V^*$ decay 
receives QCD corrections when the Higgs couples to gluons (via a heavy top
quark loop) which then couple to the weak bosons via a massless quark loop. 
This may be relevant for Higgs searches in the mass regions where the Higgs
decays into two off-shell $W$ bosons. In pure QCD, one can evaluate the
two-loop triple gluon and quark-gluon two-loop vertices with massless quarks 
in a covariant gauge (as well as the gluon-ghost vertex).\footnote{Note that
Davydychev and Osland have studied the two-loop case where only one of the legs
is off-shell~\cite{DO}}  This is a useful input for Schwinger-Dyson studies of
confinement as well as exploring how the Ward-Slavnov-Taylor identities
generalise to the off-shell case. 

As in
Refs.~\cite{GRplanar,GRnonplanar,bhabhavertex,bonciani1,bonciani2,bonciani3,czakon}, 
we employ the differential equation technique to evaluate the MI.   By building
up from the simplest MI, we find differential equations in the external scales
for new MI where the inhomogeneous terms involve known MI.   Together with
appropriate boundary conditions, the integral can then be systematically solved
order by order in $\epsilon$. At each order we encounter one-dimensional
integrals over the terms in the result for one order lower.   These integrals
yield polylogarithms and, because of the specific kinematics of the vertex
graph with three off-shell legs, we find it necessary to extend the set of
two-dimensional harmonic polylogarithms (2-d HPL) to include quadratic factors in
the denominator (see also Ref.~\cite{bonciani3}).  Our results are therefore
presented in terms of the extended  set of 2-d HPL's.

Our paper is organised as follows.   In Section~\ref{sec:method}, we give a
brief summary of the differential equation method for solving MI while in
Section~\ref{sec:method}.  We define our notation and kinematics in
Section~\ref{sec:kinematics} before  we discuss harmonic polylogarithms (HPL) and the extended set of
2-d HPL necessary for these MI in Section~\ref{sec:HPL}. In any two-loop process, some of the graphs are
factorisable and are given as products of one-loop graphs.   Therefore, we list
the one-loop MI in Section~\ref{sec:oneloop}.   We note that expressions to all orders in
$\epsilon$ for  the one-loop triangle with off-shell legs have been provided in
terms of log-sine functions by Davydychev~\cite{allorderstriangle}. 
The generalised log-sine functions can be directly related to
Nielsen polylogarithms \cite{Kalmykov} and the all-order epsilon-expansion of
one-loop massless vertex diagrams  with three off-shell legs is given
in terms of Nielsen polylogarithms in Ref.~\cite{Kalmykov}.
 However, here we
convert these results into 2-d HPL so that they can simply be combined with the
other two-loop MI. Expressions for the two-loop MI are listed   in
Section~\ref{sec:twoloop},
ordered by the number of propagators.   Finally, our results are summarized in 
Section~\ref{sec:summary}.

\section{Using differential equations to calculate master integrals}
\label{sec:method}

\subsection{The differential equations}

As discussed in Section~\ref{sec:intro}, we use the differential equation (DE)
method first suggested in \cite{Kotikov} to evaluate the MI's.  This method was
expanded in detail in \cite{GR-DE}, and has since been used to calculate many
two-loop MI, including those with multiple off-shell legs or  internal masses
\cite{GRplanar,GRnonplanar,bhabhavertex,bonciani1,bonciani2,bonciani3,czakon}.
Here we present only a brief outline of this method, and for further details
in the use of this method to find the $\epsilon$-expansion of master integrals
we refer the reader to Refs.~\cite{GRplanar,bhabhavertex}.

The main idea of this method is to derive differential equations in external
invariants for the master integrals. These equations are then solved using
suitable boundary conditions to fix the constants of integration.

The differential equations are obtained by differentiating the MI with respect
to the external momenta. Via the chain rule, linear combinations of these
differentials are used to find the first order differential equations in the
external invariants. These equations involve tensor loop integrals with
additional powers of propagators. However, the IBP and LI identities can be
used to simplify the equations\footnote{To achieve this we have made extensive
use of the Laporta algorithm~\cite{Laporta}.   We have coded our own version
using FORM~\cite{FORM} and checked it against the Automatic Integral Reduction
package AIR~\cite{AIR}.}
 so that the differential equation is expressed
only in terms of the MI itself and combinations of simpler topologies (i.e.
integrals with less denominators).  For this reason it is sensible to apply a
'bottom-up' approach, working from simpler to more complicated topologies, so
that the MI is the only unknown in the differential equation. In other words,
 the inhomogeneous
part of the differential equation is known (or at least the relevant terms in
the $\epsilon$-expansion are known). For some topologies there is more than one
MI  leading to coupled differential equations.  However as there is some
freedom in the choice of which two-loop graphs to use as the master integrals,
it simplifies the calculation to choose the master integrals such that they
have different leading powers of $\epsilon$. In this case, the system of
differential equations decouples on expansion in $\epsilon$ (see, for example,
\cite{bhabhavertex}) .

The differential equations are exact in the space-time dimension $d$, and 
can be solved as follows. Consider the inhomogeneous differential equation for
the MI $F$ with respect to the external scale $x$,
\begin{equation}
\label{eq:basicde}
\frac{d}{dx} F(x) = A(x) F(x) + B(x) .
\end{equation}
If $H(x)$ is a solution of the homogeneous equation
\begin{equation}
\label{eq:homo}
\frac{d}{dx} H(x) = A(x) H(x) .
\end{equation}
then the full solution is given by
\begin{equation}
F(x) = H(x) \left( \int^x H^{-1}(x') B(x') dx'   +  C \right)
\end{equation}
where the constant $C$ has to be fixed from the boundary conditions. 
These solutions are generally combinations of hypergeometric functions 
which are difficult to expand in powers of $\epsilon$. 
Thus to find the $\epsilon$-expansion of the MI we must 
systematically expand each master integral $F$, and all 
$d$-dependent terms of the differential equation in powers of $\epsilon$
\begin{equation}
F=\sum_{i=-m}^{n}{f^{i} \epsilon^{i}},\qquad\qquad
A=\sum_{i=0}^{n+m}{a^{i} \epsilon^{i}} ,\qquad\qquad
B=\sum_{i=-m}^{n}{b^{i} \epsilon^{i}}
\end{equation}
where $-m$ is the lowest power of  $\epsilon$ in the expansion and 
$n$ is the highest power of $\epsilon$ needed.
It is assumed that the $a^i$ and $b^i$ are already known.
Each coefficient $f^i$ satisfies the differential equation given by,
\begin{equation}
\frac{d}{dx} f^i (x) = \sum_{j=0}^{m-i} a^{j}(x) f^{i-j}(x) + b^{i}(x) .
\end{equation}
It can be seen that the homogeneous part of all the
equations generated by the $\epsilon$-expansion is simply the homogeneous 
solution $H$ evaluated at $d=4$
\begin{equation}
 h(x)=H(x)|_{d=4}
\end{equation}
and so the solution is given by
\begin{equation}
f^{i}(x) = h(x) \left( \int^x h^{-1}(x') 
\left( \sum_{j=1}^{m-i} a^j(x')f^{i-j}(x') + b^{i}(x') \right) dx'    
+  c^{i} \right)
\end{equation}
where the constants $c^{i}$ have to be fixed from the boundary conditions 
at each
order in $\epsilon$. Note that in general each coefficient
$f^{i}$ will depend on $f^{i-1}$ so we solve the system of 
equations order by order, using repeated
integration of the lower order results. 
It is for this reason that we require that $A(x)$ has no poles in 
$\epsilon$ as then $f^{i}$ would depend on $f^{i+1}$ and
the bottom-up approach would not be valid.

\subsection{The boundary conditions}

In general, the lowest order coefficient in $\epsilon$ is determined solely
by the boundary conditions. The boundary conditions are either obtained from
the differential equation or from the master integral itself. To obtain
limits from the differential equation it is necessary to examine the singular
points in the coefficients of the differential equation. For
example, if eq.~\ref{eq:basicde} were to take the form
\begin{equation}
\frac{d}{dx} F(x) = \frac{1}{x-a} F(x) + \frac{1}{x-a}B(x)
\end{equation}
then we could multiply the whole equation by $(x-a)$ and let $x
\rightarrow a$, then we have
\begin{equation}
0=F(x)|_{x=a}+B(x)|_{x=a}
\end{equation}
giving the boundary condition on $F(x)$. To obtain boundary conditions from
the integral itself, we can use limits where the $\epsilon$ expansion is known, 
for example where
an offshell leg becomes massless. In both methods care has to be
taken that the integral has a smooth limit at the chosen point 
so as not to miss or introduce hidden singularities.

If $H(x)$ is divergent at the boundary then the constant $C$ is
already determined by the necessary condition,
\begin{equation}
   \int^x H^{-1}(x') B(x') dx'   +  C = 0  \Big|_{x=a},
\end{equation}
which can be fulfilled by choosing the boundary point as the lower
integration limit. The solution of the differential equation is
then given by
\begin{equation}
    F(x) = H(x)\tilde F(x) = H(x) \left( \int^x_{a} H^{-1}(x') B(x') dx' \right) .
\end{equation}
It can be easily shown that this function satisfies the boundary condition.

\section{Kinematics}
\label{sec:kinematics}

We consider the vertex graph with three off-shell legs such that,
\begin{equation}
0 \to p_1 + p_2 + p_3.
\end{equation}
The three kinematic scales are $p_i^2$.
It is convenient to use the dimensionless variables 
\begin{equation}
  \label{eq:2}
  x = \frac{p_1^2}{p_3^2}, \quad y = \frac{p_2^2}{p_3^2}
\end{equation}
together with the determinant of the $2 \times 2$ Gram matrix,
\begin{equation}
  \label{eq:3}
  \lambda \equiv \lambda(x,y) = \sqrt{(1-x-y)^2 - 4 xy} = \sqrt{(x-x_0)(x-x_1)}
\end{equation}
with
\begin{equation*}
  x_0 = (1-\sqrt{y})^2, \qquad  x_1 = (1+\sqrt{y})^2.
\end{equation*}
In the following the dependence of $\lambda$, $x_0$ and $x_1$ on $x$
and $y$ will be
implicitly understood.
\begin{figure}[htbp]
  \centering
  \scalebox{.6}{\includegraphics[]{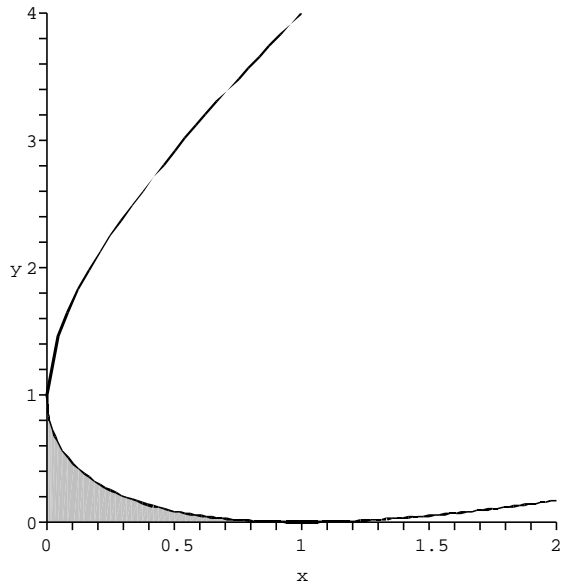}}
  \caption{The phase space for the vertex graph with three off-shell legs.
  The shaded region corresponds to the case where $p_3^2 > p_1^2,~p_2^2$.
  The solid line marks the boundary where $\lambda = 0$.}
  \label{fig:111}
\end{figure}

In this paper we choose the kinematical configuration to suit the case
of a heavy/offshell particle with momentum $p_3$ decaying into two
lighter particles with momenta $p_1,p_2$. In this
case the kinematically accessible region is given by the inequality,
\begin{equation}
\label{eq:28}
\sqrt{p_1^2} + \sqrt{p_2^2} \leq \sqrt{p_3^2},
\end{equation}
or after introduction of the dimensionless parameters
$x$ and $y$
\begin{equation}
  \sqrt{x} + \sqrt{y} \leq 1.
\end{equation}
The allowed parameter space is depicted by the shaded region in figure~\ref{fig:111}.
Other regions are simply obtained by relabelling.

\section{Harmonic polylogarithms}
\label{sec:HPL}

Solving the differential equations by repeated integration immediately suggests
that the results be given in terms of Harmonic Polylogarithms (HPL's), whose
properties are defined by repeated integration. 
The HPL's were first introduced
 in~\cite{RV} as extensions of Nielsens polylogarithms, and later extended to
 2-dimensions by~\cite{GR}.   In this section we briefly review the properties
 of one-dimensional HPL's (1-d HPL) before detailing the extension of the 2-d
 HPL's necessary to describe the vertex graph with three off-shell legs.

\subsection{1-d HPL's}

The weight, $w$, of a one-dimensional HPL, $H(\overrightarrow{b};x)$,
is the number of dimensions of the vector of parameters
$\overrightarrow{b}$. This vector, along with the argument $x$, fully
describes the HPL.

\subsubsection{1-d HPL's with $w=1$}

The weight-1 HPL's are defined as follows,
\begin{equation}
  H(a; x ) = \left \{
  \begin{array}{ll}
\int_0^x f(a,x') dx', & \qquad a\in \{1,-1\}
\\
\int_{1}^x f(a,x')dx', & \qquad a = 0
  \end{array} \right . 
\end{equation}
where,
\begin{eqnarray}
 f(1;x)&\equiv&\frac{1}{1-x}, \\
 f(0;x)&\equiv&\frac{1}{x} ,\\
 f(-1;x)&\equiv&\frac{1}{1+x}.
\end{eqnarray}
Note that $H(0;x)$ is defined differently to avoid the logarithmic
singularity at $x= 0$. Thus we have
\begin{eqnarray}
 H(1;x)&\equiv&-\ln{(1-x)} ,\\
 H(0;x)&\equiv&\ln{x}, \\
 H(-1;x)&\equiv&\ln{(1+x)}, 
\end{eqnarray}
and
\begin{equation}
\frac{\partial}{\partial x}H(a;x)=f(a;x) \qquad \mbox{with} \qquad
a\in \{+1,0,-1\}.
\end{equation}

\subsubsection{1-d HPL's with $w>1$}
The higher weight HPL's are recursively defined by,
\begin{eqnarray}
H(a,\overrightarrow{b};x)&\equiv&\int_{0}^{x}dx'f(a;x')H(\overrightarrow{b};x),\\
H(0,...,0;x)&\equiv&\frac{1}{w!}\ln^{w}{x}.
\end{eqnarray}
Note that only the HPL's with weight vectors comprising only $0$'s are
defined differently, and are 
integrated between $1$ and $x$ to avoid logarithmic
singularities. All others involve integration from $0$ to $x$.
Under differentiation, the weight is reduced by unity,
\begin{eqnarray}
\frac{\partial}{\partial
x}H(a,\overrightarrow{b};x)=f(a;x)H(\overrightarrow{b};x).
\end{eqnarray}
We also find the following relation useful:
\begin{eqnarray}
H^{n}(0;x)=n!H(\underbrace{0.....0}_{\textrm{n zeros}};x)
\end{eqnarray}

\subsection{Extension of the 2-d HPL's}
\label{sec:2dHPL}

The 2-d HPL's were introduced in~\cite{GR} as the logical extension of the 1-d
HPL's.  The common extension is the linear basis
\begin{align}
  f(a,x) &= \frac{1}{a-x} ,\\
  f(-a,x) &= \frac{1}{a+x} .
\end{align}
However, while it is possible to use this basis to evaluate the integrals
under investigation here, a more natural extension involves the square 
roots that are generated by Eq.~\ref{eq:3}.  
To this end, we extend the 2-d basis by the (quadratic) functions,
\begin{align}
  \label{eq:4}f(\lambda,x) &= \frac{1}{\lambda} ,\\
  \label{eq:5}f(x\lambda,x) &= \frac{1}{x\lambda} ,\\
  \label{eq:6}f(x_0,x) &= -\frac{1}{x-x_0}, \\
  \label{eq:7}f(x_1,x) &= -\frac{1}{x-x_1}.
\end{align}
These functions are two dimensional, with explicit dependence on $x$ and the dependence on $y$ coming from
$x_0(y)$ and $x_1(y)$.
The functions are chosen to be positive in the region,
\begin{equation*}
  0<x<1, \quad 0<y<1 ,\quad 0 < \lambda^2,
\end{equation*}
which immediately implies
\begin{equation}
  \label{eq:26}
  0 < x < x_0 = (1-\sqrt{y})^2 < 1, \quad   0 < y < y_0 =
  (1-\sqrt{x})^2 < 1 , \quad \sqrt{x} + \sqrt{y} < 1,
\end{equation}
and which corresponds to the kinetically allowed region
depicted in fig \ref{fig:111}.\footnote{It is also possible to define 
a linear basis by making the Euler transformation,
$$\lambda = 2 t + x  \quad  \Rightarrow \quad 
x = -\frac{(t - a)(t+a)}{(t-b)}$$
with
$$
  a = -\frac{1-y}{2},\quad b = -\frac{1+y}{2}
$$
such that $$
\frac{dx}{\lambda} \Rightarrow -\frac{dt}{t-b},\qquad 
\frac{dx}{\lambda^2}\Rightarrow -\frac{dt}{(t-a_1)(t-a_2)}
$$
with
$$
  a_1 = -\frac{1}{2} (1 + \sqrt{y})^2, \qquad   
  a_2 = -\frac{1}{2} (1 - \sqrt{y})^2.
$$
}

We define the extended harmonic polylogarithms in the following way,
\begin{equation}
  \label{eq:30}
  G(a,\vec w; x ,y) = \int_{x_0}^x f(a,x') G(\vec w;x',y) dx',
\end{equation}
and where the dependence on $y$ is made explicit.
Note that the lower boundary is $x = x_0$.
The only exception from this definition is 
\begin{equation}
  \label{eq:31}
  G(x_0,\ldots,x_0; x ,y) = \int_{0}^x f(x_0,x') G(x_0,\ldots,x_0;x',y) dx'.
  \end{equation}
However, HPL's of this form do not appear in the results presented in this paper.

The choice of the integration
limits is governed by the kinematic boundaries. To be able to evaluate the
extended HPL's for $x\to x_0$, the condition
$\lim_{x\to x_0} G(\vec w;x,y) = 0 $ has
to be fulfilled. This ensures that
\begin{equation}
  \label{eq:32}
  \int^x f(x_0,x') G(\vec v;x',y) dx'
\end{equation}
is finite in the limit $x\to x_0$.

Note that HPL's of the form $G(0,\vec w;x,y)$ and $G(x\lambda,\vec w;x,y)$ are
divergent in the limit that $x \to 0$.   This reflects the fact that taking the
massless limit and making the $\epsilon$-expansion does not necessarily commute.
In the cases where the $x \to 0$ limit is smooth, we find that the HPL's appear
in the combination,
\begin{equation}
\Delta G(0,\vec w;x,y) \equiv G(0,\vec w;x,y) - (1-y) G(x\lambda,\vec w;x,y)
\end{equation}
which is finite as $x \to 0$.

For the generalised HPL's of weight 1 we find,
\begin{align}
  G(0;x,y) &= \log\left(\frac{x}{x_0}\right),\\
   G(\lambda;x,y) &=   \log \left ( \frac{1-x+y-\lambda}{2\sqrt{y}}
   \right ), \\
   G(x\lambda;x,y) &= \frac{1}{1-y} \log \left (
   {\frac{(1-y)(1+x-y-\lambda)-2x}{2x\sqrt{y}} } \right),\\
   G(x_0;x,y) &= -\log\left ( \frac{x_0-x}{x_0} \right ),\\
   G(x_1;x,y) &= -\log\left ( \frac{x_1-x}{x_1-x_0} \right ).
\end{align}

In the course of solving the differential equations for the MI,
we also find integrands of the form,
\begin{equation*}
  \frac{1}{\lambda^2}, ~\frac{1}{(x-x_{0,1})\lambda} \approx  (x-x_0)^\alpha (x-x_1)^\beta .
\end{equation*}
These are not independent and can be reduced to the set of quadratic 
basis functions using the
following relation obtained via integration by parts,
\begin{eqnarray}
\lefteqn{\int (x-x_0)^\alpha (x-x_1)^\beta G(v,\vec w;x,y) dx=}\nonumber \\
&&  \frac{\alpha+\beta+2} {(\beta+1)(x_0-x_1)}   \int (x-x_0)^\alpha (x-x_1)^{\beta+1} G(v,\vec w;x,y) dx \nonumber \\ 
&& + \frac{1}{(\beta+1)(x_0-x_1)}\int (x-x_0)^{\alpha+1}(x-x_1)^{\beta+1}  f(v) G(\vec w;x,y) dx \nonumber \\ 
&& - \frac{(x-x_0)^{\alpha+1}(x-x_1)^{\beta+1} }{(\beta+1)(x_0-x_1)} G(v,\vec w ,;x,y).
\end{eqnarray}

\section{One Loop}
\label{sec:oneloop}

All one-loop master integrals contain an overall factor
\begin{equation}
\label{eq:1}
  S_D = 
  \frac{(4\pi)^\epsilon}{16\pi^2}\frac{\Gamma(1+\epsilon)\Gamma^2(1-\epsilon)}
  {\Gamma(1-2\epsilon)}.  
\end{equation}

\subsection{Two point integrals}

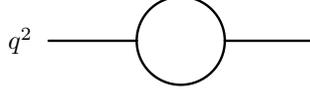
\begin{figure}[h]
\centering
\begin{fmfgraph*}(100,50)
  \fmfleft{i1}
  \fmfright{o1}
  \fmf{plain}{i1,v1}
  \fmf{plain}{v2,o1}
  \fmf{phantom}{v1,v2}
  \fmffreeze
  \fmf{plain,left}{v1,v2,v1}
  \fmfv{label=$q^2$}{i1}
\end{fmfgraph*}
\caption{The one-loop Master Integral, $\BB(q^2)$}
\label{fig:bubble}
\end{figure}
The only two point integral at one-loop level is the bubble graph
shown in figure \ref{fig:bubble} which can be easily calculated using
Feynman parameters and is given here for the sake of completeness,
\begin{equation}
\label{eq:12}  
\BB(q^2) =  i S_D (-q^2)^{-\epsilon}
  \frac{1}{\epsilon} \frac{1}{(1-2\epsilon)}.
\end{equation}

\subsection{Three point integrals}

\begin{figure}[h]
  \centering
  \begin{fmfgraph*}(100,50)
    \fmfright{i1,i2}
    \fmfleft{o1}
    \fmf{plain}{i1,v1}
    \fmf{plain}{i2,v2}
    \fmf{plain}{o1,v3}
    \fmf{plain,tension=0.4}{v1,v2,v3,v1}
    \fmfv{label=$m_1^2$,l.a=0}{i1}
    \fmfv{label=$m_2^2$,l.a=0}{i2}
    \fmfv{label=$m_3^2$}{o1}
  \end{fmfgraph*}
  \caption{The one-loop Master Integral, $F_0(m_1^2,m_2^2,m_3^2)$. }
  \label{fig:triangle}
\end{figure}
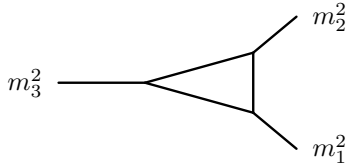
The three point integrals with less than three massive legs are
reducible to bubble integrals. The only master integral is the
triangle with three massive legs, as shown in figure~\ref{fig:triangle}.
The finite part of this diagram has been calculated in~\cite{finitetriangle}. The
$d$-dimensional result for this integral can be found in~\cite{BD,Davydychev,negdim}, where it
is given in terms of Appell functions. Davydychev calculated this
integral to all orders in terms of log-sine functions~\cite{allorderstriangle}.

In two-loop calculations this integral appears in products with
other one-loop integral. To be able to combine these integrals with
the genuine two loop integrals, it is necessary to express all
integrals in terms of the same set of functions. To achieve this we
apply the differential equation method also to the one loop triangle.

The DE for $F_0(m_1^2,m_2^2,m_3^2)$ is given by
\begin{align}
  \label{eq:4a}
\nonumber
m_1^2 \Lambda^2 \frac{\partial F_0}{\partial m_1^2}  =
&  \,\left( \left(\frac{d-4}{2}\right)\Lambda^2 + (3-d)m_1^2(m_1^2-m_2^2-m_3^2)  \right) F_0 
\\ \nonumber 
&+ (d-3)\,( m_3^2 + m_1^2 - m_2^2 ) 
{\Muserfunction{BB}}(m_3^2)\\ \nonumber  
&+ (d-3)\,(  m_1^2 + m_2^2 -m_3^2) 
\,{\Muserfunction{BB}}(m_2^2)\\  
&- 2\,(d-3)  m_1^2\,{\Muserfunction{BB}}(m_1^2) ,
\end{align}
where,
\begin{equation}
\Lambda^2 = m_1^4+m_2^4+m_3^4-2m_1^2m_2^2-2m_1^2m_3^2-2m_2^2m_3^2.
\end{equation}
The scalar triangle is completely symmetric under the interchange of the
external scales.   We choose to solve it in the configuration $m_i^2 = p_i^2$
with $p_3^2 > p_1^2,~p_2^2$.   In this case, $\Lambda^2 = p_3^4 \lambda^2$ (with
$\lambda$ given by Eq.~\ref{eq:3}) and
$m_1^2 = p_3^2 x$, $m_2^2 = p_3^2 y$.
By performing a transformation of variable $m_1^2 \equiv x\to \lambda$ it can
easily be seen that the homogeneous solution at $d=4$ is $ F_0^{hom} =
\frac{1}{\lambda}$. To fix the constant of integration it is necessary
to look for suitable boundary points. Note that the limit $x\to 0 $ is not
allowed because the one-loop with two external masses is divergent at $d=4$. The
only remaining possibility is to choose a point on the parabola given by
$\lambda = 0 $, as shown in fig \ref{fig:111}, i.e. $x\to x_{0,1} = (1\pm\sqrt{y})^2$. In
this case the homogeneous solution is divergent at the boundary and so
we apply the treatment discussed in section~\ref{sec:method}.3.
The boundary condition for $x\to x_0$, corresponding to $ m_1^2\to (m_2 - m_3)^2$ is
given by, 
\begin{equation}
  \label{eq:14}
  F_0(m_1^2,m_2^2,m_3^2)|_{m_1^2=(m_2 - m_3)^2} = 
  -  \frac{\Muserfunction{BB}({m_2}^2)}
     {\left( {m_2} - {m_3} \right)
         \,{m_3}}   + 
  \frac{\Muserfunction{BB}({\left( {m_2} - 
         {m_3} \right) }^2)}{{m_2
      }\,{m_3}} + 
  \frac{\Muserfunction{BB}({{m_3}}^2)}
   {{m_2}\,
     \left( {m_2} - {m_3} \right) }.
\end{equation}

Solving the differential equation order by order in $\epsilon$, we find the following expansion in
$\epsilon$,
\begin{equation}
\label{eq:13}
   F_0(p_1^2,p_2^2,p_3^2)=
  \begin{minipage}[c]{4cm}
\begin{fmfgraph*}(100,75)
    \fmfright{i1,i2}
    \fmfleft{o1}
    \fmf{plain}{i1,v1}
    \fmf{plain}{i2,v2}
    \fmf{plain}{o1,v3}
    \fmf{plain,tension=0.4}{v1,v2,v3,v1}
    \fmfv{label=$p_1^2$,l.a=0}{i1}
    \fmfv{label=$p_2^2$,l.a=0}{i2}
    \fmfv{label=$p_3^2$,l.a=90}{o1}
  \end{fmfgraph*}
  \end{minipage}
 = i (-p_3^2)^{-1-\epsilon} S_D \frac{1}{\lambda}
  \left ( \sum_{i=0,\ldots,2}f_0^{i}\left
  (\frac{p_1^2}{p_3^2},\frac{p_2^2}{p_3^2} \right) 
  \epsilon^i + {\cal O}(\epsilon^3)\right ),
\end{equation}
where,
\begin{align*}
  f_0^0(x,y) =&
-2\,G(\lambda ,0;x,y)
-2\,G(\lambda ;x,y)\,H(0;x_0)\\&
+G(\lambda ;x,y)\,H(0;y)
+(-1 + y)\,G(x\lambda ;x,y)\,H(0;y)
,
 \\
\\
f_0^1(x,y) =&
+2\,G(0,\lambda ,0;x,y)
+2\,G(\lambda ,0,0;x,y)
+2\,G(x_0,\lambda ,0;x,y)\\&
+2\,G(x_1,\lambda ,0;x,y)
+2\,G(\lambda ;x,y)\,H(0,0;x_0)
-\left( G(\lambda ;x,y)\,H(0,0;y) \right) \\&
+(1 - y)\,G(x\lambda ;x,y)\,H(0,0;y)
+2\,G(0,\lambda ;x,y)\,H(0;x_0)
-\left( G(0,\lambda ;x,y)\,H(0;y) \right) \\&
+(1 - y)\,G(0,x\lambda ;x,y)\,H(0;y)
+2\,G(\lambda ,0;x,y)\,H(0;x_0)
+2\,G(x_0,\lambda ;x,y)\,H(0;x_0)\\&
-\left( G(x_0,\lambda ;x,y)\,H(0;y) \right) 
+(1 - y)\,G(x_0,x\lambda ;x,y)\,H(0;y)
+2\,G(x_1,\lambda ;x,y)\,H(0;x_0)\\&
-\left( G(x_1,\lambda ;x,y)\,H(0;y) \right) 
+(1 - y)\,G(x_1,x\lambda ;x,y)\,H(0;y)
,
\\
\\
f_0^2(x,y) =&
-2\,G(0,0,\lambda ,0;x,y)
-2\,G(0,\lambda ,0,0;x,y)
-2\,G(0,x_0,\lambda ,0;x,y)\\&
-2\,G(0,x_1,\lambda ,0;x,y)
-2\,G(\lambda ,0,0,0;x,y)
-2\,G(x_0,0,\lambda ,0;x,y)\\&
-2\,G(x_0,\lambda ,0,0;x,y)
-2\,G(x_0,x_0,\lambda ,0;x,y)
-2\,G(x_0,x_1,\lambda ,0;x,y)\\&
-2\,G(x_1,0,\lambda ,0;x,y)
-2\,G(x_1,\lambda ,0,0;x,y)
-2\,G(x_1,x_0,\lambda ,0;x,y)\\&
-2\,G(x_1,x_1,\lambda ,0;x,y)
-2\,G(\lambda ;x,y)\,H(0,0,0;x_0)
+G(\lambda ;x,y)\,H(0,0,0;y)\\&
+(-1 + y)\,G(x\lambda ;x,y)\,H(0,0,0;y)
-2\,G(0,\lambda ;x,y)\,H(0,0;x_0)
+G(0,\lambda ;x,y)\,H(0,0;y)\\&
+(-1 + y)\,G(0,x\lambda ;x,y)\,H(0,0;y)
-2\,G(\lambda ,0;x,y)\,H(0,0;x_0)
-2\,G(x_0,\lambda ;x,y)\,H(0,0;x_0)\\&
+G(x_0,\lambda ;x,y)\,H(0,0;y)
+(-1 + y)\,G(x_0,x\lambda ;x,y)\,H(0,0;y)
-2\,G(x_1,\lambda ;x,y)\,H(0,0;x_0)\\&
+G(x_1,\lambda ;x,y)\,H(0,0;y)
+(-1 + y)\,G(x_1,x\lambda ;x,y)\,H(0,0;y)
-2\,G(0,0,\lambda ;x,y)\,H(0;x_0)\\&
+G(0,0,\lambda ;x,y)\,H(0;y)
+(-1 + y)\,G(0,0,x\lambda ;x,y)\,H(0;y)
-2\,G(0,\lambda ,0;x,y)\,H(0;x_0)\\&
-2\,G(0,x_0,\lambda ;x,y)\,H(0;x_0)
+G(0,x_0,\lambda ;x,y)\,H(0;y)
+(-1 + y)\,G(0,x_0,x\lambda ;x,y)\,H(0;y)\\&
-2\,G(0,x_1,\lambda ;x,y)\,H(0;x_0)
+G(0,x_1,\lambda ;x,y)\,H(0;y)
+(-1 + y)\,G(0,x_1,x\lambda ;x,y)\,H(0;y)\\&
-2\,G(\lambda ,0,0;x,y)\,H(0;x_0)
-2\,G(x_0,0,\lambda ;x,y)\,H(0;x_0)
+G(x_0,0,\lambda ;x,y)\,H(0;y)\\&
+(-1 + y)\,G(x_0,0,x\lambda ;x,y)\,H(0;y)
-2\,G(x_0,\lambda ,0;x,y)\,H(0;x_0)
-2\,G(x_0,x_0,\lambda ;x,y)\,H(0;x_0)\\&
+G(x_0,x_0,\lambda ;x,y)\,H(0;y)
+(-1 + y)\,G(x_0,x_0,x\lambda ;x,y)\,H(0;y)
-2\,G(x_0,x_1,\lambda ;x,y)\,H(0;x_0)\\&
+G(x_0,x_1,\lambda ;x,y)\,H(0;y)
+(-1 + y)\,G(x_0,x_1,x\lambda ;x,y)\,H(0;y)
-2\,G(x_1,0,\lambda ;x,y)\,H(0;x_0)\\&
+G(x_1,0,\lambda ;x,y)\,H(0;y)
+(-1 + y)\,G(x_1,0,x\lambda ;x,y)\,H(0;y)
-2\,G(x_1,\lambda ,0;x,y)\,H(0;x_0)\\&
-2\,G(x_1,x_0,\lambda ;x,y)\,H(0;x_0)
+G(x_1,x_0,\lambda ;x,y)\,H(0;y)
+(-1 + y)\,G(x_1,x_0,x\lambda ;x,y)\,H(0;y)\\&
-2\,G(x_1,x_1,\lambda ;x,y)\,H(0;x_0)
+G(x_1,x_1,\lambda ;x,y)\,H(0;y)
+(-1 + y)\,G(x_1,x_1,x\lambda ;x,y)\,H(0;y)
 .
\end{align*}
We have checked that the $\epsilon^0$ term agrees with the 
results of Ref.~\cite{finitetriangle} while the 
${\cal O}(\epsilon)$ term numerically agrees
with that given in Ref.~\cite{allorderstriangle}.

\section{Two Loop}
\label{sec:twoloop}

\subsection{Two point integrals}

There are only two two-point integrals, both of which can be obtained by
repeated one-loop integration and are widely available in the literature.
We quote them here for the sake of completeness.

\begin{figure}[ht]
  \centering
  \begin{fmfgraph*}(100,50)
    \fmfleft{i1}
    \fmfright{o1}
    \fmf{plain}{i1,v1}
    \fmf{plain}{v2,o1}
    \fmf{plain,tension=1}{v1,v2}
    \fmf{plain,left,tension=0}{v1,v2,v1}
    \fmfv{label=$q^2$}{i1}
  \end{fmfgraph*}
  \caption{The two-loop Master Integral, $\SSS(q^2)$.}
  \label{fig:sunset}
\end{figure}

The three-propagator sunset graph $\SSS(q^2)$ is shown in Fig.~\ref{fig:sunset} and is
given by,
\begin{eqnarray}
\label{eq:16}
  \SSS(q^2) &=&
S_D^2 (-q^2)^{1-2\epsilon}  
 \frac{\Gamma(2\epsilon-1)\Gamma(1-2\epsilon)^2}
{\Gamma(3-3\epsilon)\Gamma(1-\epsilon)\Gamma(1+\epsilon)^2}(-q^2)^{1-2\epsilon}  \nonumber \\
&=& S_D^2 (-q^2)^{1-2\epsilon} \left[ -\frac{1}{4\epsilon}
  -\frac{13}{8} -\frac{115}{16}\epsilon - \left ( \frac{865}{32} -
  \frac{3}{2}\zeta (3)\right )\epsilon^2 +{\cal O}(\epsilon^3)\right].
\end{eqnarray}

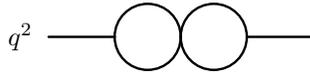
\begin{figure}[h]
  \centering
  \label{fig:gl}
  \begin{fmfgraph*}(100,50)
    \fmfleft{i1}
    \fmfright{o1}
    \fmf{phantom}{i1,v1,v2,v3,o1}
    \fmffreeze
    \fmf{plain}{i1,v1}
    \fmf{plain}{o1,v3}
    \fmf{plain,left}{v1,v2,v3,v2,v1}
    \fmfv{label=$q^2$}{i1}
  \end{fmfgraph*}
  \caption{The two-loop Master Integral, $\GL(q^2)$.}
\end{figure}

We denote the four-propagator graph shown in Fig.~\ref{fig:gl} 
by $\GL(q^2)$.   It is given by,
\begin{eqnarray}
\label{eq:17}
  \GL(q^2) = \BB(q^2)^2 &=&
  -(-q^2)^{-2\epsilon} S_D^2 \frac{1}{\epsilon^2(1-2\epsilon)^2}\nonumber \\
&=&  
  -(-q^2)^{-2\epsilon} S_D^2 \frac{1}{\epsilon^2} \left(1 + 4 \epsilon
  + 12 \epsilon^2 
  + 32 \epsilon^3 + 80 \epsilon^4 +{\cal O}(\epsilon^5) \right).
\end{eqnarray}

\subsection{Three point integrals}

\subsubsection{Master Integrals with four propagators}

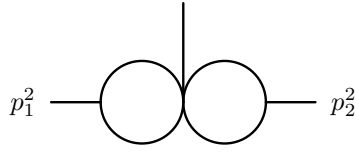
\begin{figure}[h]
\centering
\begin{fmfgraph*}(100,75)
  \fmfleft{i1}
  \fmfright{o1}
  \fmftop{t1}
  \fmf{plain}{i1,v1}
  \fmf{plain}{o1,v2}
  \fmf{plain,left=1,tension=.3}{v1,v3,v2,v3,v1}
  \fmf{plain,tension=0}{t1,v3}
  \fmfv{label=$p_1^2$,l.a=180}{i1}
  \fmfv{label=$p_2^2$,l.a=0}{o1}
\end{fmfgraph*}
\label{fig:1111}
\caption{The two-loop Master Integral, $\TGL(p_1^2,p_2^2)$}
\end{figure}

The simplest graph with four propagators is denoted by $\TGL$ and
is shown in figure~\ref{fig:1111} and is simply
the product of two bubbles,
\begin{equation}  
\TGL(p_1^2,p_2^2)=\BB(p_1^2)\BB(p_2^2)=
-(-p_1^2)^{-\epsilon}
(-p_2^2)^{-\epsilon}S_D^2\frac{1}{\epsilon^2(1-2\epsilon)^2}.
\end{equation}

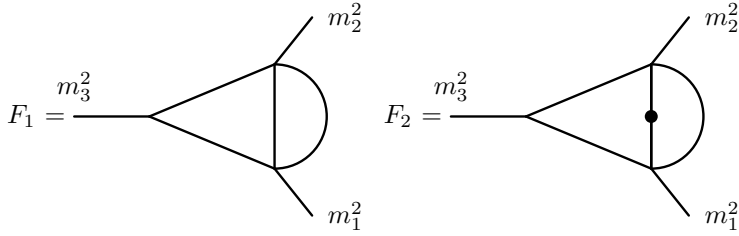
\begin{figure}[h]
\centering
$F_1$ =  
\begin{minipage}[c]{4cm}
\begin{fmfgraph*}(100,75)
  \fmfleft{i1}
  \fmfright{o1,o2}
  \fmf{plain}{i1,v1}
  \fmf{plain,tension=.3}{v1,v2,v3,v1}
  \fmf{plain}{v2,o1}
  \fmf{plain}{v3,o2}
  \fmffreeze
  \fmf{plain,left=1}{v3,v2}
  \fmfv{label=$m_3^2$,l.a=90}{i1}
  \fmfv{label=$m_1^2$,l.a=0}{o1}
  \fmfv{label=$m_2^2$,l.a=0}{o2}
\end{fmfgraph*}
\end{minipage}
$F_2$ = \begin{minipage}[c]{4cm}
\begin{fmfgraph*}(100,75)
  \fmfleft{i1}
  \fmfright{o1,o2}
  \fmf{plain}{i1,v1}
  \fmf{plain,tension=.3}{v1,v2,v3,v1}
  \fmf{plain}{v2,o1}
  \fmf{plain}{v3,o2}
  \fmffreeze
  \fmf{plain,left=1}{v3,v2}
  \fmf{plain}{v2,v4,v3}
  \fmfdot{v4}
  \fmfv{label=$m_3^2$,l.a=90}{i1}
  \fmfv{label=$m_1^2$,l.a=0}{o1}
  \fmfv{label=$m_2^2$,l.a=0}{o2}
\end{fmfgraph*}
\end{minipage}
\label{fig:f1f2}
\caption{The two-loop Master Integrals, $F_1(m_1^2,m_2^2,m_3^2)$, 
$F_2(m_1^2,m_2^2,m_3^2)$}
\end{figure}

There are two genuine two loop integrals with four propagators. These
are denoted by $F_1$ and $F_2$ and shown in figure~\ref{fig:f1f2} The
$\epsilon$ expansion of these integrals is given to order $\epsilon^0$
in~\cite{finitetwoloop}.  The corresponding two scale diagrams, when one of the
legs becomes massless, are given in~\cite{GRplanar} to order
$\epsilon^2$ in terms of HPL's.

We obtain two coupled differential equations for $F_1 $ and
$F_2$,
\begin{align}
\frac{\partial}{\partial m_1^2} &F_1(m_1^2,m_2^2,m_3^2) = \frac{\left( d-4 \right) \,
     m_2^2\,
     \left( m_1^2 - m_2^2 + 
       m_3^2 \right) }{2\,
     \left( d-3 \right) \,{\Mvariable{\Lambda}}^2} F_2\, - 
  \frac{\left( d-4 \right) \,
     \left( -m_1^2 + m_2^2 + 
       m_3^2 \right) }{2\,
     {\Mvariable{\Lambda}}^2}F_1\, \nonumber\\&+ 
  \frac{\left( 3\,d -8\right) \,
     \left( -m_1^2 - m_2^2 + 
       m_3^2 \right) \,
     }{2\,
     {\Mvariable{\Lambda}}^2\,m_1^2} \Muserfunction{SS}(m_1^2)+ 
  \frac{\left( 3\,d -8\right) }{{\Mvariable{
       \Lambda}}^2}\,
     \Muserfunction{SS}(m_2^2),\\
\nonumber\\
\frac{\partial}{\partial m_1^2} &F_2(m_1^2,m_2^2,m_3^2) =\frac{-\left( \left( d-3\right) \,(3\,d -10 )\,
              \left( m_1^2 - m_2^2 + 
         m_3^2 \right)  \right) }{2\,
     {\Mvariable{\Lambda}}^2\,m_1^2}F_1\, \nonumber\\&- 
  \frac{1 }{2\,{\Mvariable{\Lambda}}^2\,
     m_1^2}\,\bigg((3d-10)m_1^2(m_1^2-m_2^2-m_3^2)-2(d-4)\Lambda^2)
       \bigg)F_2\nonumber\\&+ 
  \frac{\left(d-3 \right) \,
     \left(  3\,d -8\right) \,(3\,d -10 )}{\left( d-4\right) \,{\Mvariable{\Lambda}}^2\,
     m_1^2} \,
     \Muserfunction{SS}(m_1^2)\nonumber\\&- 
  \frac{\left(  d-3 \right) \,
     \left(  3\,d -8\right) \,(3\,d -10 )\,
     \left( m_1^2 + m_2^2 - 
       m_3^2 \right) }{2\,
     \left( d-4 \right) \,{\Mvariable{\Lambda}}^2\,
     m_1^2\,m_2^2}\,
     \Muserfunction{SS}(m_2^2).
\end{align}
The homogeneous solutions at $d=4$ are found to be
\begin{equation*}
   F_1^{hom} = 1 , \qquad   F_2^{hom} = \frac{1}{\lambda}.
\end{equation*}
Finding suitable boundary conditions for these coupled equations is more difficult than for the one
loop triangle. Only $F_1$ has a smooth limit for $x\to 0 $ 
($F_2 $ develops additional poles in $\epsilon$ in that limit). 
On the other hand, taking the limit  $\lambda \to
0 $ in either differential equation only provides a single 
relation between the two integrals. Therefore we 
use the limit $x\to 0$ to fix the integration
constant of $ F_1$, 
matching the integral in this limit to the two-scale result given in
\cite{GRplanar}.  
Then we take the limit $\lambda \to 0 $ (i.e. $x\to x_0$) of 
$ F_1$ and use the limiting relation obtained from the differential equation
to find $ F_2$ in the limit $\lambda \to 0 $, thereby fixing the
integration constant of $ F_2$. 

The limiting relation in the limit $x\to x_0$ is,
\begin{equation}
  \label{eq:19}
\begin{split}
  F_2(m_1^2,m_2^2,m_3^2)|_{m_1^2=(m_2-m_3)^2} =& -\left( \frac{\left(  d -3 \right)
     \,F_1(m_1^2,m_2^2,m_3^2)|_{m_1^2=(m_2-m_3)^2}} 
     {m_2\,m_3} \right)  \\&+ 
  \frac{\left(  d-3 \right) \,
     \left( 3\,d -8 \right) \,
     \Muserfunction{SS}({m_2}^2)}{\left(  
       d -4 \right) \,{m_2}^2\,
     \left( m_2 - m_3 \right) \,
     m_3} - 
  \frac{\left( d-3 \right) \,
     \left(  3\,d -8\right) \,
     \Muserfunction{SS}(m_3^2)}{\left(  
       d-4 \right) \,m_2\,m_3^2\,
     \left( m_2 - m_3 \right) }.
 \end{split}
\end{equation}
Without the second boundary constraint on $F_1$ as $x \to 0$,
this would only be sufficient to yield an independent
determination of $f_1^{-2}$ only.

The expansions in $\epsilon$ are needed for two separate configurations.  
First, when the largest scale $p_3^2$ is situated opposite the bubble
(corresponding to $m_1^2 = p_3^2$) and second, when it lies adjacent to the
bubble ($m_2^2 = p_3^2$).   For the first momentum configuration,
we find that the expansions are as follows,
\begin{equation*}
  F_1^a(p_1^2,p_2^2,p_3^2)=
  \begin{minipage}[c]{4cm}
\begin{fmfgraph*}(100,75)
  \fmfleft{i1}
  \fmfright{o1,o2}
  \fmf{plain}{i1,v1}
  \fmf{plain,tension=.3}{v1,v2,v3,v1}
  \fmf{plain}{v2,o1}
  \fmf{plain}{v3,o2}
  \fmffreeze
  \fmf{plain,left=1}{v3,v2}
  \fmfv{label=$p_3^2$,l.a=90}{i1}
  \fmfv{label=$p_1^2$,l.a=0}{o1}
  \fmfv{label=$p_2^2$,l.a=0}{o2}
\end{fmfgraph*}
  \end{minipage}
 = S_D^2 (-p_3^2)^{-2\epsilon} \left ( \sum_{i=-2,\ldots,1} f_1^i\left
  (\frac{p_1^2}{p_3^2},\frac{p_2^2}{p_3^2} \right) \epsilon^i +
  {\cal O}(\epsilon^2)\right ),
\end{equation*}
where,
\begin{align*}
  f_1^{-2}(x,y) =&  - \frac{1}{2}
,
\\
\\
  f_1^{-1}(x,y) =& - \frac{5}{2} 
,
\\
\\
f_1^0(x,y) =& \frac{-57 - {\pi }^2}{6}
-H(1,0;y)
+\frac{-1 + x + y}{\lambda}\,G(\lambda ,0;x,y)
+G(\lambda ,0;0,y)\\&
-\frac{1}{2} \,\Delta G(0;0,y)\,H(0;y)
+\frac{1}{2}\,G(0;x,y)\,H(0;y)\\&
+\frac{-\left( \left( -1 + y \right) \,\left( -1 + x + y \right)  \right) }{2\,\lambda}\,G(x\lambda ;x,y)\,H(0;y)\\&
+G(\lambda ;0,y)\,H(0;x_0)
-\frac{1}{2} \,G(\lambda ;0,y)\,H(0;y)\\&
+\frac{-\left( -1 + x + y \right) }{2\,\lambda}\,G(\lambda ;x,y)\,H(0;y)
+\frac{-1 + x + y}{\lambda}\,G(\lambda ;x,y)\,H(0;x_0)
,
\\
\\
f_1^1(x,y) =&
-\left( \frac{65}{2} \right)  - \frac{5\,{\pi }^2}{6} + 2\,\zeta_3
+\frac{-{\pi }^2}{12}\,\Delta G(0;0,y)
-\Delta G(0,\lambda ,0;0,y)\\&
+\frac{{\pi }^2}{12}\,G(0;x,y)
+\frac{{\pi }^2}{12}\,G(\lambda ;0,y)
+\frac{{\pi }^2\,\left( -1 + x + y \right) }{12\,\lambda}\,G(\lambda ;x,y)\\&
+\frac{-\left( {\pi }^2\,\left( -1 + y \right) \,\left( -1 + x + y \right)  \right) }{12\,\lambda}\,G(x\lambda ;x,y)
+5\,G(\lambda ,0;0,y)\\&
+\frac{5\,\left( -1 + x + y \right) }{\lambda}\,G(\lambda ,0;x,y)
+\frac{-3\,\left( -1 + x + y \right) }{2\,\lambda}\,G(0,\lambda ,0;x,y)\\&
-G(\lambda ,0,0;0,y)
+\frac{-2\,\left( -1 + x + y \right) }{\lambda}\,G(\lambda ,0,0;x,y)\\&
-\frac{1}{2} \,G(\lambda ,0,\lambda ;0,y)
+\frac{-1 + y}{2}\,G(\lambda ,0,x\lambda ;0,y)
+\frac{1}{2}\,G(\lambda ,\lambda ,0;0,y)\\&
-\frac{3}{2} \,G(\lambda ,\lambda ,0;x,y)
+\frac{-1 + y}{2}\,G(\lambda ,x\lambda ,0;0,y)
-G(x_0,\lambda ,0;0,y)\\&
-\left( \frac{-1 + x + y}{\lambda} \right) \,G(x_0,\lambda ,0;x,y)
-G(x_1,\lambda ,0;0,y)\\&
-\left( \frac{-1 + x + y}{\lambda} \right) \,G(x_1,\lambda ,0;x,y)
+\frac{3\,\left( -1 + y \right) }{2}\,G(x\lambda ,\lambda ,0;x,y)\\&
+\frac{-{\pi }^2}{6}\,H(1;y)
-5\,H(1,0;y)
+2\,H(1,0,0;y)
-H(1,1,0;y)\\&
-\frac{5}{2} \,\Delta G(0;0,y)\,H(0;y)
+\Delta G(0;0,y)\,H(0,0;y)\\&
-\frac{1}{2} \,\Delta G(0;0,y)\,H(1,0;y)
-\frac{1}{4} \,\Delta G(0,0;0,y)\,H(0;y)\\&
-\Delta G(0,\lambda ;0,y)\,H(0;x_0) 
+\frac{3}{4}\,\Delta G(0,\lambda ;0,y)\,H(0;y)\\&
+\frac{-1 + y}{4}\,\Delta G(0,x\lambda ;0,y)\,H(0;y)
-\frac{1}{2} \,G(0;x,y)\,G(\lambda ,0;0,y)\\&
+\frac{5}{2}\,G(0;x,y)\,H(0;y)
-G(0;x,y)\,H(0,0;y) 
+\frac{1}{2}\,G(0;x,y)\,H(1,0;y)\\&
+5\,G(\lambda ;0,y)\,H(0;x_0)
-\frac{5}{2} \,G(\lambda ;0,y)\,H(0;y)
-2\,G(\lambda ;0,y)\,H(0,0;x_0)\\&
+G(\lambda ;0,y)\,H(0,0;y)
+\frac{1}{2}\,G(\lambda ;0,y)\,H(1,0;y)\\&
+\frac{-\left( -1 + x + y \right) }{2\,\lambda}\,G(\lambda ;x,y)\,G(\lambda ,0;0,y)
+\frac{5\,\left( -1 + x + y \right) }{\lambda}\,G(\lambda ;x,y)\,H(0;x_0)\\&
+\frac{-5\,\left( -1 + x + y \right) }{2\,\lambda}\,G(\lambda ;x,y)\,H(0;y)
+\frac{-2\,\left( -1 + x + y \right) }{\lambda}\,G(\lambda ;x,y)\,H(0,0;x_0)\\&
+\frac{-1 + x + y}{\lambda}\,G(\lambda ;x,y)\,H(0,0;y)
+\frac{-1 + x + y}{2\,\lambda}\,G(\lambda ;x,y)\,H(1,0;y)\\&
+\frac{\left( -1 + y \right) \,\left( -1 + x + y \right) }{2\,\lambda}\,G(x\lambda ;x,y)\,G(\lambda ,0;0,y)\\&
+\frac{-5\,\left( -1 + y \right) \,\left( -1 + x + y \right) }{2\,\lambda}\,G(x\lambda ;x,y)\,H(0;y)\\&
+\frac{\left( -1 + y \right) \,\left( -1 + x + y \right) }{\lambda}\,G(x\lambda ;x,y)\,H(0,0;y)\\&
+\frac{-\left( \left( -1 + y \right) \,\left( -1 + x + y \right)  \right) }{2\,\lambda}\,G(x\lambda ;x,y)\,H(1,0;y)\\&
-\frac{1}{4} \,G(0,0;x,y)\,H(0;y)
+\frac{-3\,\left( -1 + x + y \right) }{2\,\lambda}\,G(0,\lambda ;x,y)\,H(0;x_0)\\&
+\frac{3\,\left( -1 + x + y \right) }{4\,\lambda}\,G(0,\lambda ;x,y)\,H(0;y)\\&
+\frac{3\,\left( -1 + y \right) \,\left( -1 + x + y \right) }{4\,\lambda}\,G(0,x\lambda ;x,y)\,H(0;y)\\&
-\frac{3}{2} \,G(\lambda ,0;0,y)\,H(0;x_0)
-\frac{1}{4} \,G(\lambda ,0;0,y)\,H(0;y)\\&
+\frac{-2\,\left( -1 + x + y \right) }{\lambda}\,G(\lambda ,0;x,y)\,H(0;x_0)\\&
+\frac{-\left( -1 + x + y \right) }{4\,\lambda}\,G(\lambda ,0;x,y)\,H(0;y)\\&
+\frac{1}{2}\,G(\lambda ,\lambda ;0,y)\,H(0;x_0)
-\frac{1}{4} \,G(\lambda ,\lambda ;0,y)\,H(0;y)\\&
-\frac{3}{2} \,G(\lambda ,\lambda ;x,y)\,H(0;x_0)
+\frac{3}{4}\,G(\lambda ,\lambda ;x,y)\,H(0;y)\\&
+\frac{-1 + y}{2}\,G(\lambda ,x\lambda ;0,y)\,H(0;x_0)
+\frac{-3\,\left( -1 + y \right) }{4}\,G(\lambda ,x\lambda ;0,y)\,H(0;y)\\&
+\frac{3\,\left( -1 + y \right) }{4}\,G(\lambda ,x\lambda ;x,y)\,H(0;y)
-\left( G(x_0,\lambda ;0,y)\,H(0;x_0) \right) \\&
+\frac{1}{2}\,G(x_0,\lambda ;0,y)\,H(0;y)
-\left( \frac{-1 + x + y}{\lambda} \right) \,G(x_0,\lambda ;x,y)\,H(0;x_0)\\&
+\frac{-1 + x + y}{2\,\lambda}\,G(x_0,\lambda ;x,y)\,H(0;y)
+\frac{-1 + y}{2}\,G(x_0,x\lambda ;0,y)\,H(0;y)\\&
+\frac{\left( -1 + y \right) \,\left( -1 + x + y \right) }{2\,\lambda}\,G(x_0,x\lambda ;x,y)\,H(0;y)
-\left( G(x_1,\lambda ;0,y)\,H(0;x_0) \right) \\&
+\frac{1}{2}\,G(x_1,\lambda ;0,y)\,H(0;y)
-\left( \frac{-1 + x + y}{\lambda} \right) \,G(x_1,\lambda ;x,y)\,H(0;x_0)\\&
+\frac{-1 + x + y}{2\,\lambda}\,G(x_1,\lambda ;x,y)\,H(0;y)
+\frac{-1 + y}{2}\,G(x_1,x\lambda ;0,y)\,H(0;y)\\&
+\frac{\left( -1 + y \right) \,\left( -1 + x + y \right) }{2\,\lambda}\,G(x_1,x\lambda ;x,y)\,H(0;y)\\&
+\frac{\left( -1 + y \right) \,\left( -1 + x + y \right) }{4\,\lambda}\,G(x\lambda ,0;x,y)\,H(0;y)\\&
+\frac{3\,\left( -1 + y \right) }{2}\,G(x\lambda ,\lambda ;x,y)\,H(0;x_0)\\&
+\frac{-3\,\left( -1 + y \right) }{4}\,G(x\lambda ,\lambda ;x,y)\,H(0;y)
+\frac{-3\,{\left( -1 + y \right) }^2}{4}\,G(x\lambda ,x\lambda ;x,y)\,H(0;y)\\&
+\frac{1}{4}\,\Delta G(0;0,y)\,G(0;x,y)\,H(0;y)
+\frac{-1 + x + y}{4\,\lambda}\,\Delta G(0;0,y)\,G(\lambda ;x,y)\,H(0;y)\\&
+\frac{-\left( \left( -1 + y \right) \,\left( -1 + x + y \right)  \right) }{4\,\lambda}\,\Delta G(0;0,y)\,G(x\lambda ;x,y)\,H(0;y)\\&
-\frac{1}{2} \,G(0;x,y)\,G(\lambda ;0,y)\,H(0;x_0)
+\frac{1}{4}\,G(0;x,y)\,G(\lambda ;0,y)\,H(0;y)\\&
+\frac{-\left( -1 + x + y \right) }{2\,\lambda}\,G(\lambda ;0,y)\,G(\lambda ;x,y)\,H(0;x_0)\\&
+\frac{-1 + x + y}{4\,\lambda}\,G(\lambda ;0,y)\,G(\lambda ;x,y)\,H(0;y)\\&
+\frac{\left( -1 + y \right) \,\left( -1 + x + y \right) }{2\,\lambda}\,G(\lambda ;0,y)\,G(x\lambda ;x,y)\,H(0;x_0)\\&
+\frac{-\left( \left( -1 + y \right) \,\left( -1 + x + y \right)  \right)
}{4\,\lambda}\,G(\lambda ;0,y)\,G(x\lambda ;x,y)\,H(0;y),\\&

\end{align*}
and,
\begin{equation*}
  F_2^a(p_1^2,p_2^2,p_3^2) = 
  \begin{minipage}{4cm}
\begin{fmfgraph*}(100,75)
  \fmfleft{i1}
  \fmfright{o1,o2}
  \fmf{plain}{i1,v1}
  \fmf{plain,tension=.3}{v1,v2,v3,v1}
  \fmf{plain}{v2,o1}
  \fmf{plain}{v3,o2}
  \fmffreeze
  \fmf{plain,left=1}{v3,v2}
  \fmf{plain}{v2,v4,v3}
  \fmfdot{v4}
  \fmfv{label=$p_3^2$,l.a=90}{i1}
  \fmfv{label=$p_1^2$,l.a=0}{o1}
  \fmfv{label=$p_2^2$,l.a=0}{o2}
\end{fmfgraph*}
\end{minipage}
=
S_D^2 (-p_3^2)^{-1-2\epsilon} \frac{1}{\lambda}\left (
  \sum_{i=-1,\ldots,1} f_2^i\left
  (\frac{p_1^2}{p_3^2},\frac{p_2^2}{p_3^2} \right) \epsilon^i + 
  {\cal O}(\epsilon^2)\right ),
\end{equation*}
where,
\begin{align*}
  f_2^{-1}(x,y) =& -2\,G(\lambda ,0;x,y)
-2\,G(\lambda ;x,y)\,H(0;x_0)\\&
+G(\lambda ;x,y)\,H(0;y)
+(-1 + y)\,G(x\lambda ;x,y)\,H(0;y),\\&

 \\
f_2^{0}(x,y) = & +\frac{-{\pi }^2}{6}\,G(\lambda ;x,y)
+\frac{{\pi }^2\,\left( -1 + y \right) }{6}\,G(x\lambda ;x,y)
+3\,G(0,\lambda ,0;x,y)\\&
+4\,G(\lambda ,0,0;x,y)
+2\,G(x_0,\lambda ,0;x,y)
+2\,G(x_1,\lambda ,0;x,y)\\&
+G(\lambda ;x,y)\,G(\lambda ,0;0,y)
+4\,G(\lambda ;x,y)\,H(0,0;x_0)
-2\,G(\lambda ;x,y)\,H(0,0;y)\\&
-\left( G(\lambda ;x,y)\,H(1,0;y) \right) 
+(1 - y)\,G(x\lambda ;x,y)\,G(\lambda ,0;0,y)\\&
+(2 - 2\,y)\,G(x\lambda ;x,y)\,H(0,0;y)
+(-1 + y)\,G(x\lambda ;x,y)\,H(1,0;y)\\&
+3\,G(0,\lambda ;x,y)\,H(0;x_0)
-\frac{3}{2} \,G(0,\lambda ;x,y)\,H(0;y)\\&
+\frac{-3\,\left( -1 + y \right) }{2}\,G(0,x\lambda ;x,y)\,H(0;y)
+4\,G(\lambda ,0;x,y)\,H(0;x_0)\\&
+\frac{1}{2}\,G(\lambda ,0;x,y)\,H(0;y)
+2\,G(x_0,\lambda ;x,y)\,H(0;x_0)
-\left( G(x_0,\lambda ;x,y)\,H(0;y) \right) \\&
+(1 - y)\,G(x_0,x\lambda ;x,y)\,H(0;y)
+2\,G(x_1,\lambda ;x,y)\,H(0;x_0)
-\left( G(x_1,\lambda ;x,y)\,H(0;y) \right) \\&
+(1 - y)\,G(x_1,x\lambda ;x,y)\,H(0;y)
+\frac{1 - y}{2}\,G(x\lambda ,0;x,y)\,H(0;y)\\&
-\frac{1}{2} \,\Delta G(0;0,y)\,G(\lambda ;x,y)\,H(0;y)
+\frac{-1 + y}{2}\,\Delta G(0;0,y)\,G(x\lambda ;x,y)\,H(0;y)\\&
+G(\lambda ;0,y)\,G(\lambda ;x,y)\,H(0;x_0)
-\frac{1}{2} \,G(\lambda ;0,y)\,G(\lambda ;x,y)\,H(0;y)\\&
+(1 - y)\,G(\lambda ;0,y)\,G(x\lambda ;x,y)\,H(0;x_0)
+\frac{-1 + y}{2}\,G(\lambda ;0,y)\,G(x\lambda ;x,y)\,H(0;y),\\&
 
\\
f_2^{1}(x,y) = & 
-\zeta_3\,G(\lambda ;x,y)
+\left( -1 + y \right) \,\zeta_3\,G(x\lambda ;x,y)
+\frac{{\pi }^2}{4}\,G(0,\lambda ;x,y)\\&
+\frac{-\left( {\pi }^2\,\left( -1 + y \right)  \right) }{4}\,G(0,x\lambda ;x,y)
+\frac{{\pi }^2}{12}\,G(\lambda ,0;x,y)
+\frac{{\pi }^2}{6}\,G(x_0,\lambda ;x,y)\\&
+\frac{-\left( {\pi }^2\,\left( -1 + y \right)  \right) }{6}\,G(x_0,x\lambda ;x,y)
+\frac{{\pi }^2}{6}\,G(x_1,\lambda ;x,y)\\&
+\frac{-\left( {\pi }^2\,\left( -1 + y \right)  \right) }{6}\,G(x_1,x\lambda ;x,y)
+\frac{-\left( {\pi }^2\,\left( -1 + y \right)  \right) }{12}\,G(x\lambda ,0;x,y)\\&
-\frac{9}{2} \,G(0,0,\lambda ,0;x,y)
-6\,G(0,\lambda ,0,0;x,y)
-3\,G(0,x_0,\lambda ,0;x,y)\\&
-3\,G(0,x_1,\lambda ,0;x,y)
-8\,G(\lambda ,0,0,0;x,y)
-\frac{3}{2} \,G(\lambda ,\lambda ,\lambda ,0;x,y)\\&
+\frac{3\,\left( -1 + y \right) }{2}\,G(\lambda ,x\lambda ,\lambda ,0;x,y)
-3\,G(x_0,0,\lambda ,0;x,y)
-4\,G(x_0,\lambda ,0,0;x,y)\\&
-2\,G(x_0,x_0,\lambda ,0;x,y)
-2\,G(x_0,x_1,\lambda ,0;x,y)
-3\,G(x_1,0,\lambda ,0;x,y)\\&
-4\,G(x_1,\lambda ,0,0;x,y)
-2\,G(x_1,x_0,\lambda ,0;x,y)
-2\,G(x_1,x_1,\lambda ,0;x,y)\\&
+\frac{3\,\left( -1 + y \right) }{2}\,G(x\lambda ,\lambda ,\lambda ,0;x,y)
+\frac{-3\,{\left( -1 + y \right) }^2}{2}\,G(x\lambda ,x\lambda ,\lambda ,0;x,y)\\&
+\frac{-{\pi }^2}{12}\,\Delta G(0;0,y)\,G(\lambda ;x,y)
+\frac{{\pi }^2\,\left( -1 + y \right) }{12}\,\Delta G(0;0,y)\,G(x\lambda ;x,y)\\&
-\left( \Delta G(0,\lambda ,0;0,y)\,G(\lambda ;x,y) \right) 
+(-1 + y)\,\Delta G(0,\lambda ,0;0,y)\,G(x\lambda ;x,y)\\&
+\frac{{\pi }^2}{12}\,G(\lambda ;0,y)\,G(\lambda ;x,y)
+\frac{-\left( {\pi }^2\,\left( -1 + y \right)  \right) }{12}\,G(\lambda ;0,y)\,G(x\lambda ;x,y)\\&
-\left( G(\lambda ;x,y)\,G(\lambda ,0,0;0,y) \right) 
-\frac{1}{2} \,G(\lambda ;x,y)\,G(\lambda ,0,\lambda ;0,y)\\&
+\frac{-1 + y}{2}\,G(\lambda ;x,y)\,G(\lambda ,0,x\lambda ;0,y)
+\frac{1}{2}\,G(\lambda ;x,y)\,G(\lambda ,\lambda ,0;0,y)\\&
+\frac{-1 + y}{2}\,G(\lambda ;x,y)\,G(\lambda ,x\lambda ,0;0,y)
-\left( G(\lambda ;x,y)\,G(x_0,\lambda ,0;0,y) \right) \\&
-\left( G(\lambda ;x,y)\,G(x_1,\lambda ,0;0,y) \right) 
+\frac{-{\pi }^2}{6}\,G(\lambda ;x,y)\,H(1;y)\\&
-8\,G(\lambda ;x,y)\,H(0,0,0;x_0)
+4\,G(\lambda ;x,y)\,H(0,0,0;y)\\&
+2\,G(\lambda ;x,y)\,H(1,0,0;y)
-\left( G(\lambda ;x,y)\,H(1,1,0;y) \right) \\&
+(-1 + y)\,G(x\lambda ;x,y)\,G(\lambda ,0,0;0,y)
+\frac{-1 + y}{2}\,G(x\lambda ;x,y)\,G(\lambda ,0,\lambda ;0,y)\\&
+\frac{-{\left( -1 + y \right) }^2}{2}\,G(x\lambda ;x,y)\,G(\lambda ,0,x\lambda ;0,y)
+\frac{1 - y}{2}\,G(x\lambda ;x,y)\,G(\lambda ,\lambda ,0;0,y)\\&
+\frac{-{\left( -1 + y \right) }^2}{2}\,G(x\lambda ;x,y)\,G(\lambda ,x\lambda ,0;0,y)
+(-1 + y)\,G(x\lambda ;x,y)\,G(x_0,\lambda ,0;0,y)\\&
+(-1 + y)\,G(x\lambda ;x,y)\,G(x_1,\lambda ,0;0,y)
+\frac{{\pi }^2\,\left( -1 + y \right) }{6}\,G(x\lambda ;x,y)\,H(1;y)\\&
+4\,\left( -1 + y \right) \,G(x\lambda ;x,y)\,H(0,0,0;y)
+(2 - 2\,y)\,G(x\lambda ;x,y)\,H(1,0,0;y)\\&
+(-1 + y)\,G(x\lambda ;x,y)\,H(1,1,0;y)
-\frac{3}{2} \,G(0,\lambda ;x,y)\,G(\lambda ,0;0,y)\\&
-6\,G(0,\lambda ;x,y)\,H(0,0;x_0)
+3\,G(0,\lambda ;x,y)\,H(0,0;y)\\&
+\frac{3}{2}\,G(0,\lambda ;x,y)\,H(1,0;y)
+\frac{3\,\left( -1 + y \right) }{2}\,G(0,x\lambda ;x,y)\,G(\lambda ,0;0,y)\\&
+3\,\left( -1 + y \right) \,G(0,x\lambda ;x,y)\,H(0,0;y)
+\frac{-3\,\left( -1 + y \right) }{2}\,G(0,x\lambda ;x,y)\,H(1,0;y)\\&
-\frac{1}{2} \,G(\lambda ,0;0,y)\,G(\lambda ,0;x,y)
-\left( G(\lambda ,0;0,y)\,G(x_0,\lambda ;x,y) \right) \\&
+(-1 + y)\,G(\lambda ,0;0,y)\,G(x_0,x\lambda ;x,y)
-\left( G(\lambda ,0;0,y)\,G(x_1,\lambda ;x,y) \right) \\&
+(-1 + y)\,G(\lambda ,0;0,y)\,G(x_1,x\lambda ;x,y)
+\frac{-1 + y}{2}\,G(\lambda ,0;0,y)\,G(x\lambda ,0;x,y)\\&
-8\,G(\lambda ,0;x,y)\,H(0,0;x_0)
-\left( G(\lambda ,0;x,y)\,H(0,0;y) \right) \\&
+\frac{1}{2}\,G(\lambda ,0;x,y)\,H(1,0;y)
-4\,G(x_0,\lambda ;x,y)\,H(0,0;x_0)\\&
+2\,G(x_0,\lambda ;x,y)\,H(0,0;y)
+G(x_0,\lambda ;x,y)\,H(1,0;y)\\&
+2\,\left( -1 + y \right) \,G(x_0,x\lambda ;x,y)\,H(0,0;y)
+(1 - y)\,G(x_0,x\lambda ;x,y)\,H(1,0;y)\\&
-4\,G(x_1,\lambda ;x,y)\,H(0,0;x_0)
+2\,G(x_1,\lambda ;x,y)\,H(0,0;y)
+G(x_1,\lambda ;x,y)\,H(1,0;y)\\&
+2\,\left( -1 + y \right) \,G(x_1,x\lambda ;x,y)\,H(0,0;y)
+(1 - y)\,G(x_1,x\lambda ;x,y)\,H(1,0;y)\\&
+(-1 + y)\,G(x\lambda ,0;x,y)\,H(0,0;y)
+\frac{1 - y}{2}\,G(x\lambda ,0;x,y)\,H(1,0;y)\\&
-\frac{9}{2} \,G(0,0,\lambda ;x,y)\,H(0;x_0)
+\frac{9}{4}\,G(0,0,\lambda ;x,y)\,H(0;y)\\&
+\frac{9\,\left( -1 + y \right) }{4}\,G(0,0,x\lambda ;x,y)\,H(0;y)
-6\,G(0,\lambda ,0;x,y)\,H(0;x_0)\\&
-\frac{3}{4} \,G(0,\lambda ,0;x,y)\,H(0;y)
-3\,G(0,x_0,\lambda ;x,y)\,H(0;x_0)\\&
+\frac{3}{2}\,G(0,x_0,\lambda ;x,y)\,H(0;y)
+\frac{3\,\left( -1 + y \right) }{2}\,G(0,x_0,x\lambda ;x,y)\,H(0;y)\\&
-3\,G(0,x_1,\lambda ;x,y)\,H(0;x_0)
+\frac{3}{2}\,G(0,x_1,\lambda ;x,y)\,H(0;y)\\&
+\frac{3\,\left( -1 + y \right) }{2}\,G(0,x_1,x\lambda ;x,y)\,H(0;y)
+\frac{3\,\left( -1 + y \right) }{4}\,G(0,x\lambda ,0;x,y)\,H(0;y)\\&
-8\,G(\lambda ,0,0;x,y)\,H(0;x_0)
-\frac{1}{4} \,G(\lambda ,0,0;x,y)\,H(0;y)\\&
-\frac{3}{2} \,G(\lambda ,\lambda ,\lambda ;x,y)\,H(0;x_0)
+\frac{3}{4}\,G(\lambda ,\lambda ,\lambda ;x,y)\,H(0;y)\\&
+\frac{3\,\left( -1 + y \right) }{4}\,G(\lambda ,\lambda ,x\lambda ;x,y)\,H(0;y)
+\frac{3\,\left( -1 + y \right) }{2}\,G(\lambda ,x\lambda ,\lambda ;x,y)\,H(0;x_0)\\&
+\frac{-3\,\left( -1 + y \right) }{4}\,G(\lambda ,x\lambda ,\lambda ;x,y)\,H(0;y)
+\frac{-3\,{\left( -1 + y \right) }^2}{4}\,G(\lambda ,x\lambda ,x\lambda ;x,y)\,H(0;y)\\&
-3\,G(x_0,0,\lambda ;x,y)\,H(0;x_0)
+\frac{3}{2}\,G(x_0,0,\lambda ;x,y)\,H(0;y)\\&
+\frac{3\,\left( -1 + y \right) }{2}\,G(x_0,0,x\lambda ;x,y)\,H(0;y)
-4\,G(x_0,\lambda ,0;x,y)\,H(0;x_0)\\&
-\frac{1}{2} \,G(x_0,\lambda ,0;x,y)\,H(0;y)
-2\,G(x_0,x_0,\lambda ;x,y)\,H(0;x_0)\\&
+G(x_0,x_0,\lambda ;x,y)\,H(0;y)
+(-1 + y)\,G(x_0,x_0,x\lambda ;x,y)\,H(0;y)\\&
-2\,G(x_0,x_1,\lambda ;x,y)\,H(0;x_0)
+G(x_0,x_1,\lambda ;x,y)\,H(0;y)\\&
+(-1 + y)\,G(x_0,x_1,x\lambda ;x,y)\,H(0;y)
+\frac{-1 + y}{2}\,G(x_0,x\lambda ,0;x,y)\,H(0;y)\\&
-3\,G(x_1,0,\lambda ;x,y)\,H(0;x_0)
+\frac{3}{2}\,G(x_1,0,\lambda ;x,y)\,H(0;y)\\&
+\frac{3\,\left( -1 + y \right) }{2}\,G(x_1,0,x\lambda ;x,y)\,H(0;y)
-4\,G(x_1,\lambda ,0;x,y)\,H(0;x_0)\\&
-\frac{1}{2} \,G(x_1,\lambda ,0;x,y)\,H(0;y)
-2\,G(x_1,x_0,\lambda ;x,y)\,H(0;x_0)\\&
+G(x_1,x_0,\lambda ;x,y)\,H(0;y)
+(-1 + y)\,G(x_1,x_0,x\lambda ;x,y)\,H(0;y)\\&
-2\,G(x_1,x_1,\lambda ;x,y)\,H(0;x_0)\\&
+G(x_1,x_1,\lambda ;x,y)\,H(0;y)
+(-1 + y)\,G(x_1,x_1,x\lambda ;x,y)\,H(0;y)\\&
+\frac{-1 + y}{2}\,G(x_1,x\lambda ,0;x,y)\,H(0;y)
+\frac{-1 + y}{4}\,G(x\lambda ,0,0;x,y)\,H(0;y)\\&
+\frac{3\,\left( -1 + y \right) }{2}\,G(x\lambda ,\lambda ,\lambda ;x,y)\,H(0;x_0)
+\frac{-3\,\left( -1 + y \right) }{4}\,G(x\lambda ,\lambda ,\lambda ;x,y)\,H(0;y)\\&
+\frac{-3\,{\left( -1 + y \right) }^2}{4}\,G(x\lambda ,\lambda ,x\lambda ;x,y)\,H(0;y)
+\frac{-3\,{\left( -1 + y \right) }^2}{2}\,G(x\lambda ,x\lambda ,\lambda ;x,y)\,H(0;x_0)\\&
+\frac{3\,{\left( -1 + y \right) }^2}{4}\,G(x\lambda ,x\lambda ,\lambda ;x,y)\,H(0;y)
+\frac{3\,{\left( -1 + y \right) }^3}{4}\,G(x\lambda ,x\lambda ,x\lambda ;x,y)\,H(0;y)\\&
+\Delta G(0;0,y)\,G(\lambda ;x,y)\,H(0,0;y)
-\frac{1}{2} \,\Delta G(0;0,y)\,G(\lambda ;x,y)\,H(1,0;y)\\&
+(1 - y)\,\Delta G(0;0,y)\,G(x\lambda ;x,y)\,H(0,0;y)
+\frac{-1 + y}{2}\,\Delta G(0;0,y)\,G(x\lambda ;x,y)\,H(1,0;y)\\&
+\frac{3}{4}\,\Delta G(0;0,y)\,G(0,\lambda ;x,y)\,H(0;y)
+\frac{-3\,\left( -1 + y \right) }{4}\,\Delta G(0;0,y)\,G(0,x\lambda ;x,y)\,H(0;y)\\&
+\frac{1}{4}\,\Delta G(0;0,y)\,G(\lambda ,0;x,y)\,H(0;y)
+\frac{1}{2}\,\Delta G(0;0,y)\,G(x_0,\lambda ;x,y)\,H(0;y)\\&
+\frac{1 - y}{2}\,\Delta G(0;0,y)\,G(x_0,x\lambda ;x,y)\,H(0;y)
+\frac{1}{2}\,\Delta G(0;0,y)\,G(x_1,\lambda ;x,y)\,H(0;y)\\&
+\frac{1 - y}{2}\,\Delta G(0;0,y)\,G(x_1,x\lambda ;x,y)\,H(0;y)
+\frac{1 - y}{4}\,\Delta G(0;0,y)\,G(x\lambda ,0;x,y)\,H(0;y)\\&
-\frac{1}{4} \,\Delta G(0,0;0,y)\,G(\lambda ;x,y)\,H(0;y)
+\frac{-1 + y}{4}\,\Delta G(0,0;0,y)\,G(x\lambda ;x,y)\,H(0;y)\\&
-\left( \Delta G(0,\lambda ;0,y)\,G(\lambda ;x,y)\,H(0;x_0) \right) 
+\frac{3}{4}\,\Delta G(0,\lambda ;0,y)\,G(\lambda ;x,y)\,H(0;y)\\&
+(-1 + y)\,\Delta G(0,\lambda ;0,y)\,G(x\lambda ;x,y)\,H(0;x_0)
+\frac{-3\,\left( -1 + y \right) }{4}\,\Delta G(0,\lambda ;0,y)\,G(x\lambda ;x,y)\,H(0;y)\\&
+\frac{-1 + y}{4}\,\Delta G(0,x\lambda ;0,y)\,G(\lambda ;x,y)\,H(0;y)
+\frac{-{\left( -1 + y \right) }^2}{4}\,\Delta G(0,x\lambda ;0,y)\,G(x\lambda ;x,y)\,H(0;y)\\&
-2\,G(\lambda ;0,y)\,G(\lambda ;x,y)\,H(0,0;x_0)
+G(\lambda ;0,y)\,G(\lambda ;x,y)\,H(0,0;y)\\&
+\frac{1}{2}\,G(\lambda ;0,y)\,G(\lambda ;x,y)\,H(1,0;y)
+2\,\left( -1 + y \right) \,G(\lambda ;0,y)\,G(x\lambda ;x,y)\,H(0,0;x_0)\\&
+(1 - y)\,G(\lambda ;0,y)\,G(x\lambda ;x,y)\,H(0,0;y)
+\frac{1 - y}{2}\,G(\lambda ;0,y)\,G(x\lambda ;x,y)\,H(1,0;y)\\&
-\frac{3}{2} \,G(\lambda ;0,y)\,G(0,\lambda ;x,y)\,H(0;x_0)
+\frac{3}{4}\,G(\lambda ;0,y)\,G(0,\lambda ;x,y)\,H(0;y)\\&
+\frac{3\,\left( -1 + y \right) }{2}\,G(\lambda ;0,y)\,G(0,x\lambda ;x,y)\,H(0;x_0)
+\frac{-3\,\left( -1 + y \right) }{4}\,G(\lambda ;0,y)\,G(0,x\lambda ;x,y)\,H(0;y)\\&
-\frac{1}{2} \,G(\lambda ;0,y)\,G(\lambda ,0;x,y)\,H(0;x_0)
+\frac{1}{4}\,G(\lambda ;0,y)\,G(\lambda ,0;x,y)\,H(0;y)\\&
-\left( G(\lambda ;0,y)\,G(x_0,\lambda ;x,y)\,H(0;x_0) \right) 
+\frac{1}{2}\,G(\lambda ;0,y)\,G(x_0,\lambda ;x,y)\,H(0;y)\\&
+(-1 + y)\,G(\lambda ;0,y)\,G(x_0,x\lambda ;x,y)\,H(0;x_0)
+\frac{1 - y}{2}\,G(\lambda ;0,y)\,G(x_0,x\lambda ;x,y)\,H(0;y)\\&
-\left( G(\lambda ;0,y)\,G(x_1,\lambda ;x,y)\,H(0;x_0) \right) 
+\frac{1}{2}\,G(\lambda ;0,y)\,G(x_1,\lambda ;x,y)\,H(0;y)\\&
+(-1 + y)\,G(\lambda ;0,y)\,G(x_1,x\lambda ;x,y)\,H(0;x_0)
+\frac{1 - y}{2}\,G(\lambda ;0,y)\,G(x_1,x\lambda ;x,y)\,H(0;y)\\&
+\frac{-1 + y}{2}\,G(\lambda ;0,y)\,G(x\lambda ,0;x,y)\,H(0;x_0)
+\frac{1 - y}{4}\,G(\lambda ;0,y)\,G(x\lambda ,0;x,y)\,H(0;y)\\&
-\frac{3}{2} \,G(\lambda ;x,y)\,G(\lambda ,0;0,y)\,H(0;x_0)
-\frac{1}{4} \,G(\lambda ;x,y)\,G(\lambda ,0;0,y)\,H(0;y)\\&
+\frac{1}{2}\,G(\lambda ;x,y)\,G(\lambda ,\lambda ;0,y)\,H(0;x_0)
-\frac{1}{4} \,G(\lambda ;x,y)\,G(\lambda ,\lambda ;0,y)\,H(0;y)\\&
+\frac{-1 + y}{2}\,G(\lambda ;x,y)\,G(\lambda ,x\lambda ;0,y)\,H(0;x_0)
+\frac{-3\,\left( -1 + y \right) }{4}\,G(\lambda ;x,y)\,G(\lambda ,x\lambda ;0,y)\,H(0;y)\\&
-\left( G(\lambda ;x,y)\,G(x_0,\lambda ;0,y)\,H(0;x_0) \right) 
+\frac{1}{2}\,G(\lambda ;x,y)\,G(x_0,\lambda ;0,y)\,H(0;y)\\&
+\frac{-1 + y}{2}\,G(\lambda ;x,y)\,G(x_0,x\lambda ;0,y)\,H(0;y)
-\left( G(\lambda ;x,y)\,G(x_1,\lambda ;0,y)\,H(0;x_0) \right) \\&
+\frac{1}{2}\,G(\lambda ;x,y)\,G(x_1,\lambda ;0,y)\,H(0;y)
+\frac{-1 + y}{2}\,G(\lambda ;x,y)\,G(x_1,x\lambda ;0,y)\,H(0;y)\\&
+\frac{3\,\left( -1 + y \right) }{2}\,G(x\lambda ;x,y)\,G(\lambda ,0;0,y)\,H(0;x_0)
+\frac{-1 + y}{4}\,G(x\lambda ;x,y)\,G(\lambda ,0;0,y)\,H(0;y)\\&
+\frac{1 - y}{2}\,G(x\lambda ;x,y)\,G(\lambda ,\lambda ;0,y)\,H(0;x_0)
+\frac{-1 + y}{4}\,G(x\lambda ;x,y)\,G(\lambda ,\lambda ;0,y)\,H(0;y)\\&
+\frac{-{\left( -1 + y \right) }^2}{2}\,G(x\lambda ;x,y)\,G(\lambda ,x\lambda ;0,y)\,H(0;x_0)
+\frac{3\,{\left( -1 + y \right) }^2}{4}\,G(x\lambda ;x,y)\,G(\lambda ,x\lambda ;0,y)\,H(0;y)\\&
+(-1 + y)\,G(x\lambda ;x,y)\,G(x_0,\lambda ;0,y)\,H(0;x_0)
+\frac{1 - y}{2}\,G(x\lambda ;x,y)\,G(x_0,\lambda ;0,y)\,H(0;y)\\&
+\frac{-{\left( -1 + y \right) }^2}{2}\,G(x\lambda ;x,y)\,G(x_0,x\lambda ;0,y)\,H(0;y)
+(-1 + y)\,G(x\lambda ;x,y)\,G(x_1,\lambda ;0,y)\,H(0;x_0)\\&
+\frac{1 - y}{2}\,G(x\lambda ;x,y)\,G(x_1,\lambda ;0,y)\,H(0;y)
+\frac{-{\left( -1 + y \right) }^2}{2}\,G(x\lambda ;x,y)\,G(x_1,x\lambda ;0,y)\,H(0;y)
. 
\end{align*}

The second momentum configuration is defined by,
\begin{equation*}
\label{eq:20}
  F_1^b(p_1^2,p_2^2,p_3^2)=
  \begin{minipage}[c]{4cm}
\begin{fmfgraph*}(100,75)
  \fmfleft{i1}
  \fmfright{o1,o2}
  \fmf{plain}{i1,v1}
  \fmf{plain,tension=.3}{v1,v2,v3,v1}
  \fmf{plain}{v2,o1}
  \fmf{plain}{v3,o2}
  \fmffreeze
  \fmf{plain,left=1}{v3,v2}
  \fmfv{label=$p_2^2$,l.a=90}{i1}
  \fmfv{label=$p_1^2$,l.a=0}{o1}
  \fmfv{label=$p_3^2$,l.a=0}{o2}
\end{fmfgraph*}
  \end{minipage}
 = S_D^2 (-p_3^2)^{-2\epsilon} \left ( \sum_{i=-2,\ldots,1} f_1^i\left
  (\frac{p_1^2}{p_3^2},\frac{p_2^2}{p_3^2} \right) \epsilon^i +
  {\cal O}(\epsilon^2)\right ),
\end{equation*}
\begin{equation*}
\label{eq:21}
  F_2^b(p_1^2,p_2^2,p_3^2) = \begin{minipage}{4cm}
\begin{fmfgraph*}(100,75)
  \fmfleft{i1}
  \fmfright{o1,o2}
  \fmf{plain}{i1,v1}
  \fmf{plain,tension=.3}{v1,v2,v3,v1}
  \fmf{plain}{v2,o1}
  \fmf{plain}{v3,o2}
  \fmffreeze
  \fmf{plain,left=1}{v3,v2}
  \fmf{plain}{v2,v4,v3}
  \fmfdot{v4}
  \fmfv{label=$p_2^2$,l.a=90}{i1}
  \fmfv{label=$p_1^2$,l.a=0}{o1}
  \fmfv{label=$p_3^2$,l.a=0}{o2}
\end{fmfgraph*}
\end{minipage}
=
S_D^2 (-p_3^2)^{-1-2\epsilon} \frac{1}{\lambda}\left (
  \sum_{i=-1,\ldots,1} f_2^i\left
  (\frac{p_1^2}{p_3^2},\frac{p_2^2}{p_3^2} \right) \epsilon^i + 
  {\cal O}(\epsilon^2)\right ).
\end{equation*}
Expressions for $f_{1,2}^i$ in this momentum configuration are similarly
lengthy to those for $F_{1,2}^a$ and can be obtained in 
computer readable form from the authors.
In each case, we have checked that the leading contribution agrees with 
the results of Ref.~\cite{finitetwoloop}.
 
Note that solving the differential equation for
 the crossed triangle requires the functions
$F^a_{1,2}(x,y)$ which are symmetric in $x$ and $y$ and in addition
the functions $F^b_{1,2}(x,y)$ and $F^b_{1,2}(y,x)$. The functions
$F_{1,2}^b(y,x)$ are, of course, in principle known.  However, 
exchanging $x$ and
$y$ puts them in a form that is not
suited for further integration over $x$. 
To get them into a suitable form we therefore recalculate them directly
from the differential equations.

\subsubsection{Master Integrals with five propagators}

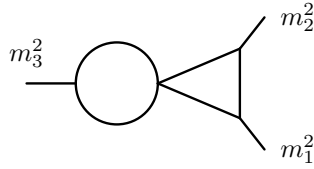
\begin{figure}[h]
  \centering
  \begin{fmfgraph*}(100,50)
    \fmfleft{i1}
    \fmfright{o1,o2}
    \fmf{plain}{i1,v1}
    \fmf{plain,left=1,tension=.3}{v1,v2,v1}
    \fmf{plain}{o1,v3}
    \fmf{plain}{o2,v4}
    \fmf{plain,tension=.3}{v2,v3,v4,v2}
  \fmfv{label=$m_3^2$,l.a=90}{i1}
  \fmfv{label=$m_1^2$,l.a=0}{o1}
  \fmfv{label=$m_2^2$,l.a=0}{o2}
  \end{fmfgraph*}
  \caption{The two-loop Master Integral, $\TB(m_1^2,m_2^2,m_3^3)$}
  \label{fig:bt}
\end{figure}
There are two master integrals with five propagators. The first is a product
of one-loop integrals and the second is a genuine two-loop
integral. The first, denoted by $\TB$, is shown in figure~\ref{fig:bt} and is given by,
\begin{equation}
\TB(p_1^2,p_2^2,p_3^3)= \BB(p_i^2) F_0(p_1^2,p_2^2,p_3^3).
\end{equation}
The $\epsilon$ expansion for this integral is straightforwardly obtained 
from eq.~\ref{eq:13} and we do not show it here.

\begin{figure}[h]
  \centering
  \begin{fmfgraph*}(100,50)
  \fmfleft{i1}
  \fmfright{o1,o2}
  \fmf{plain}{i1,v1}
  \fmf{plain}{o1,v2}
  \fmf{plain}{o2,v3}
  \fmf{plain,tension=.3}{v1,v2,v4,v3,v1}
  \fmffreeze
  \fmf{plain}{v1,v4}
  \fmfv{label=$m_1^2$,l.a=180}{i1}
  \fmfv{label=$m_2^2$,l.a=0}{o1}
  \fmfv{label=$m_3^2$,l.a=0}{o2}
  \end{fmfgraph*}
  \caption{The two-loop Master Integral, $F_3(m_1^2,m_2^2,m_3^3)$}
  \label{fig:f3}
\end{figure}
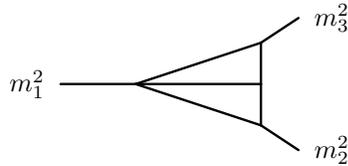

The second  irreducible two-loop diagram is denoted by $F_3$ and is shown in figure~\ref{fig:f3}.
It is finite in 4-dimensions and the leading contribution has been calculated in Ref.~\cite{finitetwoloop}.

The differential equation for $F_3$ is given by,
\begin{align}
  \frac{\partial }{\partial m_1^2} & F_3(m_1^2,m_2^2,m_3^2) = \frac{(d-3)\,
     \left( -m_1^2 + m_2^2 + 
       m_3^2 \right) }{{\Lambda}^2}\Mvariable{F_3}\,
   - \frac{2\,(3\,d -10)\,(d-3)}{
     (d-4)\,{\Lambda}^2} \,
     \Muserfunction{F_1}(m_1^2,m_2^2,m_3^2)\nonumber\\&- 
  \frac{2\,(3\,d -10)\,(d-3)}{
     (d-4)\,{\Lambda}^2}\,
     \Muserfunction{F_1}(m_3^2,m_1^2,m_2^2) + 
  \frac{\left( -m_1^2 - 
       m_2^2 + m_3^2 \right)}{\Lambda^2}  \,
     \Muserfunction{F_2}(m_1^2,m_2^2,m_3^2)
      \nonumber\\&- \frac{\left( m_1^2 - 
       m_2^2 + m_3^2 \right)}{
       {\Lambda}^2} \,
     \Muserfunction{F_2}(m_3^2,m_1^2,m_2^2) + \frac{4\,{(d-3)}^2}{(d-4)\,
     {\Lambda}^2}\,
     \Muserfunction{TGL}(m_3^2,m_2^2).
\end{align}
The solution of the homogeneous equation at $d=4$ is given by
$F_3^{hom}=\frac{1}{\lambda}$ and we therefore
take the boundary condition at $\lambda = 0$ corresponding to $x\to x_0$,
so that 
\begin{equation}
  \label{eq:22}
  \begin{split}
    F_3(m_1^2,m_2^2,m_3^2)|_{m_1^2=(m_2-m_3)^2} =& - \frac{\left(3\,d -10\right) }{\left(  d -4\right) \,
       \left( m_2 - m_3 \right) \,
       m_3}\,
       {F_1}({m_2}^2,
        {m_3}^2,
        {\left( m_2 - m_3 \right)
            }^2) \\& - 
  \frac{\left(  3\,d -10\right)}{\left( d -4 \right) \,
     \left( m_2 - m_3 \right) \,
     m_3} \,
     {F_1}({\left( m_2 - 
          m_3 \right) }^2,{m_2}^2,
      {m_3}^2) \\&- 
  \frac{m_2}{\left(  d -3\right) \,
     \left( m_2 - m_3 \right) }\,
     {F_2}({m_2}^2,
      {m_3}^2,
      {\left( m_2 - m_3 \right) }^
       2) \\&- 
  \frac{m_2}{\left( d -3\right) \,
     m_3}\,
     {F_2}({\left( m_2 - 
          m_3 \right) }^2,{m_2}^2,
      {m_3}^2) \\&+ 
  \frac{2\,\left( d -3 \right) }
     {\left( d-4 \right) \,
     \left( m_2 - m_3 \right) \,
     m_3}\,
     \Muserfunction{TGL}({\left( m_2 - 
          m_3 \right) }^2,{m_3}^2).
  \end{split}
\end{equation}
Once again, there are two distinct kinematic configurations depending
on the position of the large scale $p_3^2$. 
For the first momentum configuration, 
the first two terms in the $\epsilon$ expansion are given by,
\begin{equation*}
  F_3^a(p_1^2,p_2^2,p_3^2) =   \begin{minipage}{4cm}
  \begin{fmfgraph*}(100,75)
  \fmfleft{i1}
  \fmfright{o1,o2}
  \fmf{plain}{i1,v1}
  \fmf{plain}{o1,v2}
  \fmf{plain}{o2,v3}
  \fmf{plain,tension=.3}{v1,v2,v4,v3,v1}
  \fmffreeze
  \fmf{plain}{v1,v4}
  \fmfv{label=$p_1^2$,l.a=90}{i1}
  \fmfv{label=$p_2^2$,l.a=0}{o1}
  \fmfv{label=$p_3^2$,l.a=0}{o2}
  \end{fmfgraph*}
  \end{minipage}
= S_D^2 (-p_3^2)^{-1-2\epsilon} \frac{1}{\lambda}\left (
  \sum_{i=0,\ldots,1} f_3^i\left
  (\frac{p_1^2}{p_3^2},\frac{p_2^2}{p_3^2} \right) \epsilon^i + 
  {\cal O}(\epsilon^2)\right ),
\end{equation*}
where,
\begin{align*}
f_3^{0} =& 
-6\,\zeta_3\,G(\lambda ;x,y)
+6\,G(\lambda ,\lambda ,\lambda ,0;x,y)\\&
+\frac{-{\pi }^2}{3}\,G(\lambda ;0,y)\,G(\lambda ;x,y)
+2\,G(\lambda ;x,y)\,G(\lambda ,0,\lambda ;0,y)\\&
-2\,G(\lambda ;x,y)\,G(\lambda ,\lambda ,0;0,y)
+\frac{{\pi }^2}{3}\,G(\lambda ;x,y)\,H(0;y)\\&
+2\,G(\lambda ;x,y)\,H(0,1,0;y)
-2\,G(\lambda ;x,y)\,H(1,0,0;y)\\&
+G(\lambda ,0;x,y)\,H(0,0;y)
+6\,G(\lambda ,\lambda ,\lambda ;x,y)\,H(0;x_0)\\&
-3\,G(\lambda ,\lambda ,\lambda ;x,y)\,H(0;y)
+(3 - 3\,y)\,G(\lambda ,\lambda ,x\lambda ;x,y)\,H(0;y)\\&
- \Delta G(0;0,y)\,G(\lambda ;x,y)\,H(0,0;y) 
- \Delta G(0,\lambda ;0,y)\,G(\lambda ;x,y)\,H(0;y) \\&
- G(\lambda ;0,y)\,G(\lambda ;x,y)\,H(0,0;y) 
-2\,G(\lambda ;0,y)\,G(\lambda ;x,y)\,H(1,0;y)\\&
-2\,G(\lambda ;x,y)\,G(\lambda ,\lambda ;0,y)\,H(0;x_0)
+G(\lambda ;x,y)\,G(\lambda ,\lambda ;0,y)\,H(0;y)\\&
+2\,\left( -1 + y \right) \,G(\lambda ;x,y)\,G(\lambda ,x\lambda ;0,y)\,H(0;y)
,
\\
\\
f_3^{1} =& 
+\frac{-3\,{\pi }^4}{20}\,G(\lambda ;x,y)
-3\,\zeta_3\,G(\lambda ,0;x,y)
+6\,\zeta_3\,G(x_0,\lambda ;x,y)
+6\,\zeta_3\,G(x_1,\lambda ;x,y)\\&
+\frac{-\left( {\pi }^2\,\left( -1 + y \right)  \right) }{2}\,G(\lambda ,\lambda ,x\lambda ;x,y)
+\frac{-\left( {\pi }^2\,\left( -1 + y \right)  \right) }{2}\,G(\lambda ,x\lambda ,\lambda ;x,y)\\&
-3\,G(\lambda ,0,\lambda ,\lambda ,0;x,y)
-9\,G(\lambda ,\lambda ,0,\lambda ,0;x,y)
-12\,G(\lambda ,\lambda ,\lambda ,0,0;x,y)\\&
-6\,G(\lambda ,\lambda ,x_0,\lambda ,0;x,y)
-6\,G(\lambda ,\lambda ,x_1,\lambda ,0;x,y)
-6\,G(x_0,\lambda ,\lambda ,\lambda ,0;x,y)\\&
-6\,G(x_1,\lambda ,\lambda ,\lambda ,0;x,y)
+3\,\zeta_3\,\Delta G(0;0,y)\,G(\lambda ;x,y)
+\frac{{\pi }^2}{6}\,\Delta G(0,\lambda ;0,y)\,G(\lambda ;x,y)\\&
-3\,\Delta G(0,\lambda ,0,\lambda ;0,y)\,G(\lambda ;x,y)
-6\,\Delta G(0,\lambda ,\lambda ,0;0,y)\,G(\lambda ;x,y)
+\zeta_3\,G(\lambda ;0,y)\,G(\lambda ;x,y)\\&
+\frac{{\pi }^2}{6}\,G(\lambda ;0,y)\,G(\lambda ,0;x,y)
+\frac{{\pi }^2}{3}\,G(\lambda ;0,y)\,G(x_0,\lambda ;x,y)
+\frac{{\pi }^2}{3}\,G(\lambda ;0,y)\,G(x_1,\lambda ;x,y)\\&
+\frac{-{\pi }^2}{6}\,G(\lambda ;x,y)\,G(\lambda ,0;0,y)
+\frac{{\pi }^2\,\left( -1 + y \right) }{6}\,G(\lambda ;x,y)\,G(\lambda ,x\lambda ;0,y)
+\frac{{\pi }^2}{3}\,G(\lambda ;x,y)\,G(x_0,\lambda ;0,y)\\&
+\frac{{\pi }^2}{3}\,G(\lambda ;x,y)\,G(x_1,\lambda ;0,y)
-4\,G(\lambda ;x,y)\,G(\lambda ,0,0,\lambda ;0,y)
-\left( G(\lambda ;x,y)\,G(\lambda ,0,\lambda ,0;0,y) \right) \\&
-2\,G(\lambda ;x,y)\,G(\lambda ,\lambda ,0,0;0,y)
+(3 - 3\,y)\,G(\lambda ;x,y)\,G(\lambda ,\lambda ,0,x\lambda ;0,y)\\&
+(3 - 3\,y)\,G(\lambda ;x,y)\,G(\lambda ,\lambda ,x\lambda ,0;0,y)
+4\,G(\lambda ;x,y)\,G(\lambda ,x_0,\lambda ,0;0,y)\\&
+4\,G(\lambda ;x,y)\,G(\lambda ,x_1,\lambda ,0;0,y)
+(3 - 3\,y)\,G(\lambda ;x,y)\,G(\lambda ,x\lambda ,\lambda ,0;0,y)\\&
-2\,G(\lambda ;x,y)\,G(x_0,\lambda ,0,\lambda ;0,y)
-4\,G(\lambda ;x,y)\,G(x_0,\lambda ,\lambda ,0;0,y)
-2\,G(\lambda ;x,y)\,G(x_1,\lambda ,0,\lambda ;0,y)\\&
-4\,G(\lambda ;x,y)\,G(x_1,\lambda ,\lambda ,0;0,y)
+8\,\zeta_3\,G(\lambda ;x,y)\,H(0;y)
+6\,\zeta_3\,G(\lambda ;x,y)\,H(1;y)\\&
+\frac{-{\pi }^2}{3}\,G(\lambda ;x,y)\,H(0,0;y)
+\frac{{\pi }^2}{3}\,G(\lambda ;x,y)\,H(0,1;y)
+\frac{{\pi }^2}{3}\,G(\lambda ;x,y)\,H(1,0;y)\\&
+2\,G(\lambda ;x,y)\,H(0,0,0,0;y)
-2\,G(\lambda ;x,y)\,H(0,0,1,0;y)
-2\,G(\lambda ;x,y)\,H(0,1,0,0;y)\\&
+2\,G(\lambda ;x,y)\,H(0,1,1,0;y)
+6\,G(\lambda ;x,y)\,H(1,0,0,0;y)
+2\,G(\lambda ;x,y)\,H(1,0,1,0;y)\\&
-2\,G(\lambda ;x,y)\,H(1,1,0,0;y)
+3\,\left( -1 + y \right) \,G(\lambda ,0;0,y)\,G(\lambda ,\lambda ,x\lambda ;x,y)\\&
+3\,\left( -1 + y \right) \,G(\lambda ,0;0,y)\,G(\lambda ,x\lambda ,\lambda ;x,y)
-\left( G(\lambda ,0;x,y)\,G(\lambda ,0,\lambda ;0,y) \right) \\&
+G(\lambda ,0;x,y)\,G(\lambda ,\lambda ,0;0,y)
+\frac{-{\pi }^2}{6}\,G(\lambda ,0;x,y)\,H(0;y)
-3\,G(\lambda ,0;x,y)\,H(0,0,0;y)\\&
-\left( G(\lambda ,0;x,y)\,H(0,1,0;y) \right) 
+G(\lambda ,0;x,y)\,H(1,0,0;y)
-2\,G(x_0,\lambda ;x,y)\,G(\lambda ,0,\lambda ;0,y)\\&
+2\,G(x_0,\lambda ;x,y)\,G(\lambda ,\lambda ,0;0,y)
+\frac{-{\pi }^2}{3}\,G(x_0,\lambda ;x,y)\,H(0;y)\\&
-2\,G(x_0,\lambda ;x,y)\,H(0,1,0;y)
+2\,G(x_0,\lambda ;x,y)\,H(1,0,0;y)
-2\,G(x_1,\lambda ;x,y)\,G(\lambda ,0,\lambda ;0,y)\\&
+2\,G(x_1,\lambda ;x,y)\,G(\lambda ,\lambda ,0;0,y)
+\frac{-{\pi }^2}{3}\,G(x_1,\lambda ;x,y)\,H(0;y)
-2\,G(x_1,\lambda ;x,y)\,H(0,1,0;y)\\&
+2\,G(x_1,\lambda ;x,y)\,H(1,0,0;y)
-\frac{1}{2} \,G(\lambda ,0,0;x,y)\,H(0,0;y)
-12\,G(\lambda ,\lambda ,\lambda ;x,y)\,H(0,0;x_0)\\&
+\frac{9}{2}\,G(\lambda ,\lambda ,\lambda ;x,y)\,H(0,0;y)
+\frac{9\,\left( -1 + y \right) }{2}\,G(\lambda ,\lambda ,x\lambda ;x,y)\,H(0,0;y)\\&
+(3 - 3\,y)\,G(\lambda ,\lambda ,x\lambda ;x,y)\,H(1,0;y)
+\frac{-3\,\left( -1 + y \right) }{2}\,G(\lambda ,x\lambda ,\lambda ;x,y)\,H(0,0;y)\\&
+(3 - 3\,y)\,G(\lambda ,x\lambda ,\lambda ;x,y)\,H(1,0;y)
+\frac{-3\,{\left( -1 + y \right) }^2}{2}\,G(\lambda ,x\lambda ,x\lambda ;x,y)\,H(0,0;y)\\&
-\left( G(x_0,\lambda ,0;x,y)\,H(0,0;y) \right) 
-\left( G(x_1,\lambda ,0;x,y)\,H(0,0;y) \right) 
-3\,G(\lambda ,0,\lambda ,\lambda ;x,y)\,H(0;x_0)\\&
+\frac{3}{2}\,G(\lambda ,0,\lambda ,\lambda ;x,y)\,H(0;y)
+\frac{3\,\left( -1 + y \right) }{2}\,G(\lambda ,0,\lambda ,x\lambda ;x,y)\,H(0;y)\\&
-9\,G(\lambda ,\lambda ,0,\lambda ;x,y)\,H(0;x_0)
+\frac{9}{2}\,G(\lambda ,\lambda ,0,\lambda ;x,y)\,H(0;y)\\&
+\frac{9\,\left( -1 + y \right) }{2}\,G(\lambda ,\lambda ,0,x\lambda ;x,y)\,H(0;y)
-12\,G(\lambda ,\lambda ,\lambda ,0;x,y)\,H(0;x_0)\\&
-6\,G(\lambda ,\lambda ,x_0,\lambda ;x,y)\,H(0;x_0)
+3\,G(\lambda ,\lambda ,x_0,\lambda ;x,y)\,H(0;y)\\&
+3\,\left( -1 + y \right) \,G(\lambda ,\lambda ,x_0,x\lambda ;x,y)\,H(0;y)
-6\,G(\lambda ,\lambda ,x_1,\lambda ;x,y)\,H(0;x_0)\\&
+3\,G(\lambda ,\lambda ,x_1,\lambda ;x,y)\,H(0;y)
+3\,\left( -1 + y \right) \,G(\lambda ,\lambda ,x_1,x\lambda ;x,y)\,H(0;y)\\&
+\frac{3\,\left( -1 + y \right) }{2}\,G(\lambda ,\lambda ,x\lambda ,0;x,y)\,H(0;y)
+\frac{3\,\left( -1 + y \right) }{2}\,G(\lambda ,x\lambda ,\lambda ,0;x,y)\,H(0;y)\\&
-6\,G(x_0,\lambda ,\lambda ,\lambda ;x,y)\,H(0;x_0)
+3\,G(x_0,\lambda ,\lambda ,\lambda ;x,y)\,H(0;y)\\&
+3\,\left( -1 + y \right) \,G(x_0,\lambda ,\lambda ,x\lambda ;x,y)\,H(0;y)
-6\,G(x_1,\lambda ,\lambda ,\lambda ;x,y)\,H(0;x_0)\\&
+3\,G(x_1,\lambda ,\lambda ,\lambda ;x,y)\,H(0;y)
+3\,\left( -1 + y \right) \,G(x_1,\lambda ,\lambda ,x\lambda ;x,y)\,H(0;y)\\&
+\frac{{\pi }^2}{6}\,\Delta G(0;0,y)\,G(\lambda ;x,y)\,H(0;y)
+3\,\Delta G(0;0,y)\,G(\lambda ;x,y)\,H(0,0,0;y)\\&
+\Delta G(0;0,y)\,G(\lambda ;x,y)\,H(0,1,0;y)
-\left( \Delta G(0;0,y)\,G(\lambda ;x,y)\,H(1,0,0;y) \right) \\&
+\frac{1}{2}\,\Delta G(0;0,y)\,G(\lambda ,0;x,y)\,H(0,0;y)
+\Delta G(0;0,y)\,G(x_0,\lambda ;x,y)\,H(0,0;y)\\&
+\Delta G(0;0,y)\,G(x_1,\lambda ;x,y)\,H(0,0;y)
+\frac{-3\,\left( -1 + y \right) }{2}\,\Delta G(0;0,y)\,G(\lambda ,\lambda ,x\lambda ;x,y)\,H(0;y)\\&
+\frac{-3\,\left( -1 + y \right) }{2}\,\Delta G(0;0,y)\,G(\lambda ,x\lambda ,\lambda ;x,y)\,H(0;y)
-\frac{1}{2} \,\Delta G(0,0;0,y)\,G(\lambda ;x,y)\,H(0,0;y)\\&
-\left( \Delta G(0,\lambda ;0,y)\,G(\lambda ;x,y)\,G(\lambda ,0;0,y) \right) 
+\frac{5}{2}\,\Delta G(0,\lambda ;0,y)\,G(\lambda ;x,y)\,H(0,0;y)\\&
+\Delta G(0,\lambda ;0,y)\,G(\lambda ;x,y)\,H(1,0;y)
+\frac{1}{2}\,\Delta G(0,\lambda ;0,y)\,G(\lambda ,0;x,y)\,H(0;y)\\&
+\Delta G(0,\lambda ;0,y)\,G(x_0,\lambda ;x,y)\,H(0;y)
+\Delta G(0,\lambda ;0,y)\,G(x_1,\lambda ;x,y)\,H(0;y)\\&
+\frac{-1 + y}{2}\,\Delta G(0,x\lambda ;0,y)\,G(\lambda ;x,y)\,H(0,0;y)
+\frac{3}{2}\,\Delta G(0,0,\lambda ;0,y)\,G(\lambda ;x,y)\,H(0;y)\\&
-\frac{1}{2} \,\Delta G(0,\lambda ,0;0,y)\,G(\lambda ;x,y)\,H(0;y)
-8\,\Delta G(0,\lambda ,\lambda ;0,y)\,G(\lambda ;x,y)\,H(0;x_0)\\&
+4\,\Delta G(0,\lambda ,\lambda ;0,y)\,G(\lambda ;x,y)\,H(0;y)
+\frac{5\,\left( -1 + y \right) }{2}\,\Delta G(0,\lambda ,x\lambda ;0,y)\,G(\lambda ;x,y)\,H(0;y)\\&
+\Delta G(0,x_0,\lambda ;0,y)\,G(\lambda ;x,y)\,H(0;y)
+\Delta G(0,x_1,\lambda ;0,y)\,G(\lambda ;x,y)\,H(0;y)\\&
+\frac{5\,\left( -1 + y \right) }{2}\,\Delta G(0,x\lambda ,\lambda ;0,y)\,G(\lambda ;x,y)\,H(0;y)
+\frac{{\pi }^2}{6}\,G(\lambda ;0,y)\,G(\lambda ;x,y)\,H(0;y)\\&
+\frac{-{\pi }^2}{3}\,G(\lambda ;0,y)\,G(\lambda ;x,y)\,H(1;y)
+3\,G(\lambda ;0,y)\,G(\lambda ;x,y)\,H(0,0,0;y)\\&
+G(\lambda ;0,y)\,G(\lambda ;x,y)\,H(0,1,0;y)
+3\,G(\lambda ;0,y)\,G(\lambda ;x,y)\,H(1,0,0;y)\\&
-2\,G(\lambda ;0,y)\,G(\lambda ;x,y)\,H(1,1,0;y)
+\frac{1}{2}\,G(\lambda ;0,y)\,G(\lambda ,0;x,y)\,H(0,0;y)\\&
+G(\lambda ;0,y)\,G(\lambda ,0;x,y)\,H(1,0;y)
+G(\lambda ;0,y)\,G(x_0,\lambda ;x,y)\,H(0,0;y)\\&
+2\,G(\lambda ;0,y)\,G(x_0,\lambda ;x,y)\,H(1,0;y)
+G(\lambda ;0,y)\,G(x_1,\lambda ;x,y)\,H(0,0;y)\\&
+2\,G(\lambda ;0,y)\,G(x_1,\lambda ;x,y)\,H(1,0;y)
+3\,\left( -1 + y \right) \,G(\lambda ;0,y)\,G(\lambda ,\lambda ,x\lambda ;x,y)\,H(0;x_0)\\&
+\frac{-3\,\left( -1 + y \right) }{2}\,G(\lambda ;0,y)\,G(\lambda ,\lambda ,x\lambda ;x,y)\,H(0;y)
+3\,\left( -1 + y \right) \,G(\lambda ;0,y)\,G(\lambda ,x\lambda ,\lambda ;x,y)\,H(0;x_0)\\&
+\frac{-3\,\left( -1 + y \right) }{2}\,G(\lambda ;0,y)\,G(\lambda ,x\lambda ,\lambda ;x,y)\,H(0;y)
+G(\lambda ;x,y)\,{G(\lambda ,0;0,y)}^2\\&
+(1 - y)\,G(\lambda ;x,y)\,G(\lambda ,0;0,y)\,G(\lambda ,x\lambda ;0,y)
-2\,G(\lambda ;x,y)\,G(\lambda ,0;0,y)\,G(x_0,\lambda ;0,y)\\&
-2\,G(\lambda ;x,y)\,G(\lambda ,0;0,y)\,G(x_1,\lambda ;0,y)
-\frac{1}{2} \,G(\lambda ;x,y)\,G(\lambda ,0;0,y)\,H(0,0;y)\\&
-\left( G(\lambda ;x,y)\,G(\lambda ,0;0,y)\,H(1,0;y) \right) 
+4\,G(\lambda ;x,y)\,G(\lambda ,\lambda ;0,y)\,H(0,0;x_0)\\&
-\frac{3}{2} \,G(\lambda ;x,y)\,G(\lambda ,\lambda ;0,y)\,H(0,0;y)
+\frac{-7\,\left( -1 + y \right) }{2}\,G(\lambda ;x,y)\,G(\lambda ,x\lambda ;0,y)\,H(0,0;y)\\&
+(-1 + y)\,G(\lambda ;x,y)\,G(\lambda ,x\lambda ;0,y)\,H(1,0;y)
+G(\lambda ;x,y)\,G(x_0,\lambda ;0,y)\,H(0,0;y)\\&
+2\,G(\lambda ;x,y)\,G(x_0,\lambda ;0,y)\,H(1,0;y)
+(-1 + y)\,G(\lambda ;x,y)\,G(x_0,x\lambda ;0,y)\,H(0,0;y)\\&
+G(\lambda ;x,y)\,G(x_1,\lambda ;0,y)\,H(0,0;y)
+2\,G(\lambda ;x,y)\,G(x_1,\lambda ;0,y)\,H(1,0;y)\\&
+(-1 + y)\,G(\lambda ;x,y)\,G(x_1,x\lambda ;0,y)\,H(0,0;y)
-\left( G(\lambda ;x,y)\,G(\lambda ,0,\lambda ;0,y)\,H(0;x_0) \right) \\&
-\frac{3}{2} \,G(\lambda ;x,y)\,G(\lambda ,0,\lambda ;0,y)\,H(0;y)
+(1 - y)\,G(\lambda ;x,y)\,G(\lambda ,0,x\lambda ;0,y)\,H(0;y)\\&
+3\,G(\lambda ;x,y)\,G(\lambda ,\lambda ,0;0,y)\,H(0;x_0)
+\frac{1}{2}\,G(\lambda ;x,y)\,G(\lambda ,\lambda ,0;0,y)\,H(0;y)\\&
+(5 - 5\,y)\,G(\lambda ;x,y)\,G(\lambda ,\lambda ,x\lambda ;0,y)\,H(0;x_0)
+\frac{5\,\left( -1 + y \right) }{2}\,G(\lambda ;x,y)\,G(\lambda ,\lambda ,x\lambda ;0,y)\,H(0;y)\\&
+2\,G(\lambda ;x,y)\,G(\lambda ,x_0,\lambda ;0,y)\,H(0;x_0)
-\left( G(\lambda ;x,y)\,G(\lambda ,x_0,\lambda ;0,y)\,H(0;y) \right) \\&
+(2 - 2\,y)\,G(\lambda ;x,y)\,G(\lambda ,x_0,x\lambda ;0,y)\,H(0;y)
+2\,G(\lambda ;x,y)\,G(\lambda ,x_1,\lambda ;0,y)\,H(0;x_0)\\&
-\left( G(\lambda ;x,y)\,G(\lambda ,x_1,\lambda ;0,y)\,H(0;y) \right) 
+(2 - 2\,y)\,G(\lambda ;x,y)\,G(\lambda ,x_1,x\lambda ;0,y)\,H(0;y)\\&
+(-1 + y)\,G(\lambda ;x,y)\,G(\lambda ,x\lambda ,0;0,y)\,H(0;y)
+(5 - 5\,y)\,G(\lambda ;x,y)\,G(\lambda ,x\lambda ,\lambda ;0,y)\,H(0;x_0)\\&
+\frac{5\,\left( -1 + y \right) }{2}\,G(\lambda ;x,y)\,G(\lambda ,x\lambda ,\lambda ;0,y)\,H(0;y)
+4\,{\left( -1 + y \right) }^2\,G(\lambda ;x,y)\,G(\lambda ,x\lambda ,x\lambda ;0,y)\,H(0;y)\\&
+G(\lambda ;x,y)\,G(x_0,0,\lambda ;0,y)\,H(0;y)
-8\,G(\lambda ;x,y)\,G(x_0,\lambda ,\lambda ;0,y)\,H(0;x_0)\\&
+4\,G(\lambda ;x,y)\,G(x_0,\lambda ,\lambda ;0,y)\,H(0;y)
+2\,\left( -1 + y \right) \,G(\lambda ;x,y)\,G(x_0,\lambda ,x\lambda ;0,y)\,H(0;y)\\&
+2\,\left( -1 + y \right) \,G(\lambda ;x,y)\,G(x_0,x\lambda ,\lambda ;0,y)\,H(0;y)
+G(\lambda ;x,y)\,G(x_1,0,\lambda ;0,y)\,H(0;y)\\&
-8\,G(\lambda ;x,y)\,G(x_1,\lambda ,\lambda ;0,y)\,H(0;x_0)
+4\,G(\lambda ;x,y)\,G(x_1,\lambda ,\lambda ;0,y)\,H(0;y)\\&
+2\,\left( -1 + y \right) \,G(\lambda ;x,y)\,G(x_1,\lambda ,x\lambda ;0,y)\,H(0;y)
+2\,\left( -1 + y \right) \,G(\lambda ;x,y)\,G(x_1,x\lambda ,\lambda ;0,y)\,H(0;y)\\&
+G(\lambda ,0;x,y)\,G(\lambda ,\lambda ;0,y)\,H(0;x_0)
-\frac{1}{2} \,G(\lambda ,0;x,y)\,G(\lambda ,\lambda ;0,y)\,H(0;y)\\&
+(1 - y)\,G(\lambda ,0;x,y)\,G(\lambda ,x\lambda ;0,y)\,H(0;y)
+2\,G(\lambda ,\lambda ;0,y)\,G(x_0,\lambda ;x,y)\,H(0;x_0)\\&
-\left( G(\lambda ,\lambda ;0,y)\,G(x_0,\lambda ;x,y)\,H(0;y) \right) 
+2\,G(\lambda ,\lambda ;0,y)\,G(x_1,\lambda ;x,y)\,H(0;x_0)\\&
-\left( G(\lambda ,\lambda ;0,y)\,G(x_1,\lambda ;x,y)\,H(0;y) \right) 
+(2 - 2\,y)\,G(\lambda ,x\lambda ;0,y)\,G(x_0,\lambda ;x,y)\,H(0;y)\\&
+(2 - 2\,y)\,G(\lambda ,x\lambda ;0,y)\,G(x_1,\lambda ;x,y)\,H(0;y)
.
\end{align*}

The other momentum configuration required is defined by,
\begin{equation*}
  F_3^b(p_1^2,p_2^2,p_3^2) = \begin{minipage}{4cm}\begin{fmfgraph*}(100,75)
  \fmfleft{i1}
  \fmfright{o1,o2}
  \fmf{plain}{i1,v1}
  \fmf{plain}{o1,v2}
  \fmf{plain}{o2,v3}
  \fmf{plain,tension=.3}{v1,v2,v4,v3,v1}
  \fmffreeze
  \fmf{plain}{v1,v4}
  \fmfv{label=$p_3^2$,l.a=90}{i1}
  \fmfv{label=$p_2^2$,l.a=0}{o1}
  \fmfv{label=$p_1^2$,l.a=0}{o2}
  \end{fmfgraph*}
  \end{minipage}=
  S_D^2 (-p_3^2)^{-1-2\epsilon} \frac{1}{\lambda}\left (
  \sum_{i=0,\ldots,1} f_3^i\left
  (\frac{p_1^2}{p_3^2},\frac{p_2^2}{p_3^2} \right) \epsilon^i + 
  {\cal O}(\epsilon^2)\right ).
\end{equation*}
Expressions for $f_{3}^i$ in this momentum configuration are similarly
lengthy to those for $F_{3}^a$ and can be obtained in 
computer readable form from the authors.
In each case, we have checked that the finite contribution agrees with 
the results of Ref.~\cite{finitetwoloop}.

\subsubsection{Master Integrals with six propagators}

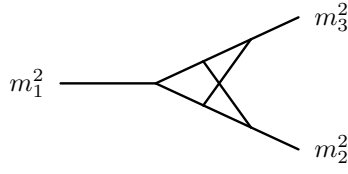
\begin{figure}[h]
  \centering
  \begin{fmfgraph*}(100,50)
  \fmfleft{i1}
  \fmfright{o1,o2}
  \fmf{plain}{i1,v1}
  \fmf{plain}{o1,v2,v3,v1,v4,v5,o2}
  \fmffreeze
  \fmf{plain}{v2,v4}
  \fmf{plain}{v3,v5}
  \fmfv{label=$m_1^2$,l.a=180}{i1}
  \fmfv{label=$m_2^2$,l.a=0}{o1}
  \fmfv{label=$m_3^2$,l.a=0}{o2}
  \end{fmfgraph*}
  \caption{The two-loop Master Integral, $F_4(m_1^2,m_2^2,m_3^3)$}
  \label{fig:f4}
\end{figure}
The only MI with 6 propagators is the crossed triangle which is denoted by $F_4$ and 
is shown in figure~\ref{fig:f4}.

This integral is fully symmetric in all three legs.
and satisfies the differential equation,
\begin{align}
  \label{eq:23}
\frac{\partial}{\partial m_1^2} F_4 =& \frac{(6-d) \,
     \left( -{m_1}^2 + {m_2}^2 + 
       {m_3}^2 \right) }{{\Lambda}^2}\, F_4  - 
  \frac{8}{{\Lambda}^2}\, F_2 ({m_1}^2,{m_2}^2,{m_3}^2) \nonumber\\&
  + \frac{4\,\left( {m_1}^2 + 
       {m_2}^2 - {m_3}^2 \right) }{
     {\Lambda}^2\,{m_1}^2} \,
      F_2 ({m_2}^2,
      {m_3}^2,{m_1}^2)- 
  \frac{4\,\left( -{m_1}^2 + 
       {m_2}^2 - {m_3}^2 \right) }{
     {\Lambda}^2\,{m_1}^2} \,
      F_2 ({m_3}^2,
      {m_2}^2,{m_1}^2)\nonumber\\&- 
  \frac{4\,\Mvariable{(d-4)}\,
      F_3 ({m_1}^2,
      {m_2}^2,{m_3}^2)}{{\Lambda}^
     2} + \frac{2\,\Mvariable{(d-4)}\,
     \left( {m_1}^2 + {m_2}^2 - 
       {m_3}^2 \right) }{
     {\Lambda}^2\,{m_1}^2} \,
      F_3 ({m_2}^2,
      {m_3}^2,{m_1}^2)\nonumber\\&- 
  \frac{2\,\Mvariable{(d-4)}\,
     \left( -{m_1}^2 + {m_2}^2 - 
       {m_3}^2 \right) }{
     {\Lambda}^2\,{m_1}^2}\,
      F_3 ({m_3}^2,
      {m_2}^2,{m_1}^2).
\end{align}
The homogeneous solution  at $d=4$ is $F_4^{hom}=\lambda^{-2}$ while 
the boundary condition at $\lambda = 0$ corresponding to $x \to x_0$ is given
by,
\begin{align}
&F_4(m_1^2,m_2^2,m_3^2)|_{m_1^2=(m_2-m_3)^2} = \frac{4\,
     \left( m_2 - m_3 \right) \,
     }{(d-6)\,m_1^2\,
     m_3}F_2(m_2^2,
      m_3^2,
      {\left( m_2 - m_3 \right) }^
       2) \nonumber\\&- 
  \frac{4\,}{(d-6)\,
     m_2\,m_3}F_2({\left( m_2 - 
          m_3 \right) }^2,m_2^2,
      m_3^2) - 
  \frac{4\,\left( m_2 - m_3 \right)
       \,}{(d-6)\,m_1^2\,
     m_2}F_2(m_3^2,
      m_2^2,
      {\left( m_2 - m_3 \right) }^
       2) \nonumber\\&+ 
  \frac{2\,(d-4)\,
     \left( m_2 - m_3 \right) \,
     }{(d-6)\,m_1^2\,
     m_3}F_3(m_2^2,
      m_3^2,
      {\left( m_2 - m_3 \right) }^
       2) - 
  \frac{2\,(d-4)\,
     }{(d-6)\,
     m_2\,m_3} F_3({\left( m_2 - 
          m_3 \right) }^2,m_2^2,
      m_3^2)\nonumber \\&- 
  \frac{2\,(d-4)\,
     \left( m_2 - m_3 \right) \,
     }{(d-6)\,m_1^2\,
     m_2}F_3(m_3^2,
      m_2^2,
      {\left( m_2 - m_3 \right) }^
       2).
 \end{align}

We find that the first two terms of the $\epsilon$-expansion are given by,
\begin{equation*}
F_4(p_1^2,p_2^2,p_3^2) = 
\begin{minipage}[c]{4cm}
\begin{fmfgraph*}(100,50)
  \fmfleft{i1}
  \fmfright{o1,o2}
  \fmf{plain}{i1,v1}
  \fmf{plain}{o1,v2,v3,v1,v4,v5,o2}
  \fmffreeze
  \fmf{plain}{v2,v4}
  \fmf{plain}{v3,v5}
  \fmfv{label=$p_1^2$,l.a=90}{i1}
  \fmfv{label=$p_2^2$,l.a=0}{o1}
  \fmfv{label=$p_3^2$,l.a=0}{o2}
\end{fmfgraph*}
\end{minipage}
 = S_D^2 (-p_3^2)^{-2\epsilon} \left ( \sum_{i=0,\ldots,1} f_1^i\left
  (\frac{p_1^2}{p_3^2},\frac{p_2^2}{p_3^2} \right) \epsilon^i +
  {\cal O}(\epsilon^2)\right ),
\end{equation*}
where,
\begin{align*}
  f_4^{0}(x,y) =&  +\frac{-8\,{\pi }^2}{3}\,G(\lambda ,\lambda ;x,y)
+\frac{-4\,{\pi }^2\,\left( -1 + y \right) }{3}\,G(\lambda ,x\lambda ;x,y)
+\frac{-4\,{\pi }^2\,\left( -1 + y \right) }{3}\,G(x\lambda ,\lambda ;x,y)\\&
-8\,G(\lambda ,0,\lambda ,0;x,y)
-16\,G(\lambda ,\lambda ,0,0;x,y)
+16\,G(\lambda ,0;0,x_0)\,G(\lambda ,\lambda ;x,y)\\&
+8\,\left( -1 + y \right) \,G(\lambda ,0;0,y)\,G(\lambda ,x\lambda ;x,y)
+8\,\left( -1 + y \right) \,G(\lambda ,0;0,y)\,G(x\lambda ,\lambda ;x,y)\\&
-16\,G(\lambda ,\lambda ;x,y)\,H(0,0;x_0)
-4\,G(\lambda ,\lambda ;x,y)\,H(0,0;y)
-16\,G(\lambda ,\lambda ;x,y)\,H(1,0;x_0)\\&
+(4 - 4\,y)\,G(\lambda ,x\lambda ;x,y)\,H(0,0;y)
+(8 - 8\,y)\,G(\lambda ,x\lambda ;x,y)\,H(1,0;y)\\&
+(4 - 4\,y)\,G(x\lambda ,\lambda ;x,y)\,H(0,0;y)
+(8 - 8\,y)\,G(x\lambda ,\lambda ;x,y)\,H(1,0;y)\\&
-4\,{\left( -1 + y \right) }^2\,G(x\lambda ,x\lambda ;x,y)\,H(0,0;y)
-8\,G(\lambda ,0,\lambda ;x,y)\,H(0;x_0)\\&
+4\,G(\lambda ,0,\lambda ;x,y)\,H(0;y)
+4\,\left( -1 + y \right) \,G(\lambda ,0,x\lambda ;x,y)\,H(0;y)\\&
-16\,G(\lambda ,\lambda ,0;x,y)\,H(0;x_0)
+8\,G(\lambda ,\lambda ,0;x,y)\,H(0;y)\\&
+4\,\left( -1 + y \right) \,G(\lambda ,x\lambda ,0;x,y)\,H(0;y)
+4\,\left( -1 + y \right) \,G(x\lambda ,\lambda ,0;x,y)\,H(0;y)\\&
-8\,\Delta G(0;0,x_0)\,G(\lambda ,\lambda ;x,y)\,H(0;x_0)
+(4 - 4\,y)\,\Delta G(0;0,y)\,G(\lambda ,x\lambda ;x,y)\,H(0;y)\\&
+(4 - 4\,y)\,\Delta G(0;0,y)\,G(x\lambda ,\lambda ;x,y)\,H(0;y)
-8\,G(\lambda ;0,x_0)\,G(\lambda ,\lambda ;x,y)\,H(0;x_0)\\&
+16\,G(\lambda ;0,x_0)\,G(\lambda ,\lambda ;x,y)\,H(0;y)
+8\,\left( -1 + y \right) \,G(\lambda ;0,y)\,G(\lambda ,x\lambda ;x,y)\,H(0;x_0)\\&
+(4 - 4\,y)\,G(\lambda ;0,y)\,G(\lambda ,x\lambda ;x,y)\,H(0;y)
+8\,\left( -1 + y \right) \,G(\lambda ;0,y)\,G(x\lambda ,\lambda ;x,y)\,H(0;x_0)\\&
+(4 - 4\,y)\,G(\lambda ;0,y)\,G(x\lambda ,\lambda ;x,y)\,H(0;y)

\\
  f_4^{1}(x,y) =& +112\,\zeta_3\,G(\lambda ,\lambda ;x,y)
+8\,\left( -1 + y \right) \,\zeta_3\,G(\lambda ,x\lambda ;x,y)
+8\,\left( -1 + y \right) \,\zeta_3\,G(x\lambda ,\lambda ;x,y)\\&
+32\,{\left( -1 + y \right) }^2\,\zeta_3\,G(x\lambda ,x\lambda ;x,y)
+\frac{4\,{\pi }^2}{3}\,G(\lambda ,0,\lambda ;x,y)
+\frac{4\,{\pi }^2\,\left( -1 + y \right) }{3}\,G(\lambda ,0,x\lambda ;x,y)\\&
+\frac{4\,{\pi }^2}{3}\,G(\lambda ,\lambda ,0;x,y)
+\frac{8\,{\pi }^2}{3}\,G(\lambda ,x_0,\lambda ;x,y)
+\frac{4\,{\pi }^2\,\left( -1 + y \right) }{3}\,G(\lambda ,x_0,x\lambda ;x,y)\\&
+\frac{8\,{\pi }^2}{3}\,G(\lambda ,x_1,\lambda ;x,y)
+\frac{4\,{\pi }^2\,\left( -1 + y \right) }{3}\,G(\lambda ,x_1,x\lambda ;x,y)
+\frac{4\,{\pi }^2\,\left( -1 + y \right) }{3}\,G(\lambda ,x\lambda ,0;x,y)\\&
+\frac{-8\,{\pi }^2}{3}\,G(x_0,\lambda ,\lambda ;x,y)
+\frac{-4\,{\pi }^2\,\left( -1 + y \right) }{3}\,G(x_0,\lambda ,x\lambda ;x,y)
+\frac{-4\,{\pi }^2\,\left( -1 + y \right) }{3}\,G(x_0,x\lambda ,\lambda ;x,y)\\&
+\frac{-8\,{\pi }^2}{3}\,G(x_1,\lambda ,\lambda ;x,y)
+\frac{-4\,{\pi }^2\,\left( -1 + y \right) }{3}\,G(x_1,\lambda ,x\lambda ;x,y)
+\frac{-4\,{\pi }^2\,\left( -1 + y \right) }{3}\,G(x_1,x\lambda ,\lambda ;x,y)\\&
+\frac{8\,{\pi }^2\,\left( -1 + y \right) }{3}\,G(x\lambda ,0,\lambda ;x,y)
+\frac{4\,{\pi }^2\,\left( -1 + y \right) }{3}\,G(x\lambda ,\lambda ,0;x,y)
+\frac{4\,{\pi }^2\,\left( -1 + y \right) }{3}\,G(x\lambda ,x_0,\lambda ;x,y)\\&
+\frac{4\,{\pi }^2\,\left( -1 + y \right) }{3}\,G(x\lambda ,x_1,\lambda ;x,y)
+\frac{-4\,{\pi }^2\,{\left( -1 + y \right) }^2}{3}\,G(x\lambda ,x\lambda ,0;x,y)
+16\,G(\lambda ,0,0,\lambda ,0;x,y)\\&
+24\,G(\lambda ,0,\lambda ,0,0;x,y)
+8\,G(\lambda ,0,x_0,\lambda ,0;x,y)
+8\,G(\lambda ,0,x_1,\lambda ,0;x,y)\\&
+48\,G(\lambda ,\lambda ,0,0,0;x,y)
-72\,G(\lambda ,\lambda ,\lambda ,\lambda ,0;x,y)
+8\,G(\lambda ,x_0,0,\lambda ,0;x,y)\\&
+16\,G(\lambda ,x_0,\lambda ,0,0;x,y)
+8\,G(\lambda ,x_1,0,\lambda ,0;x,y)
+16\,G(\lambda ,x_1,\lambda ,0,0;x,y)\\&
+24\,{\left( -1 + y \right) }^2\,G(\lambda ,x\lambda ,x\lambda ,\lambda ,0;x,y)
-8\,G(x_0,\lambda ,0,\lambda ,0;x,y)
-16\,G(x_0,\lambda ,\lambda ,0,0;x,y)\\&
-8\,G(x_1,\lambda ,0,\lambda ,0;x,y)
-16\,G(x_1,\lambda ,\lambda ,0,0;x,y)
+24\,{\left( -1 + y \right) }^2\,G(x\lambda ,\lambda ,x\lambda ,\lambda ,0;x,y)\\&
+24\,{\left( -1 + y \right) }^2\,G(x\lambda ,x\lambda ,\lambda ,\lambda ,0;x,y)
+\frac{-2\,{\pi }^2}{3}\,\Delta G(0;0,x_0)\,G(\lambda ,\lambda ;x,y)\\&
+\frac{2\,{\pi }^2\,{\left( -1 + y \right) }^2}{3}\,\Delta G(0;0,x_0)\,G(x\lambda ,x\lambda ;x,y)
+\frac{-4\,{\pi }^2\,\left( -1 + y \right) }{3}\,\Delta G(0;0,y)\,G(\lambda ,x\lambda ;x,y)\\&
+\frac{-4\,{\pi }^2\,\left( -1 + y \right) }{3}\,\Delta G(0;0,y)\,G(x\lambda ,\lambda ;x,y)
-12\,\Delta G(0,\lambda ,0;0,x_0)\,G(\lambda ,\lambda ;x,y)\\&
+12\,{\left( -1 + y \right) }^2\,\Delta G(0,\lambda ,0;0,x_0)\,G(x\lambda ,x\lambda ;x,y)
-16\,\left( -1 + y \right) \,\Delta G(0,\lambda ,0;0,y)\,G(\lambda ,x\lambda ;x,y)\\&
-16\,\left( -1 + y \right) \,\Delta G(0,\lambda ,0;0,y)\,G(x\lambda ,\lambda ;x,y)
+\frac{-2\,{\pi }^2}{3}\,G(\lambda ;0,x_0)\,G(\lambda ,\lambda ;x,y)\\&
+\frac{2\,{\pi }^2\,{\left( -1 + y \right) }^2}{3}\,G(\lambda ;0,x_0)\,G(x\lambda ,x\lambda ;x,y)
+\frac{8\,{\pi }^2}{3}\,G(\lambda ;0,y)\,G(\lambda ,\lambda ;x,y)\\&
-8\,G(\lambda ,0;0,x_0)\,G(\lambda ,0,\lambda ;x,y)
-8\,G(\lambda ,0;0,x_0)\,G(\lambda ,\lambda ,0;x,y)\\&
-16\,G(\lambda ,0;0,x_0)\,G(\lambda ,x_0,\lambda ;x,y)
-16\,G(\lambda ,0;0,x_0)\,G(\lambda ,x_1,\lambda ;x,y)\\&
+16\,G(\lambda ,0;0,x_0)\,G(x_0,\lambda ,\lambda ;x,y)
+16\,G(\lambda ,0;0,x_0)\,G(x_1,\lambda ,\lambda ;x,y)\\&
+8\,{\left( -1 + y \right) }^2\,G(\lambda ,0;0,x_0)\,G(x\lambda ,x\lambda ,0;x,y)
+(8 - 8\,y)\,G(\lambda ,0;0,y)\,G(\lambda ,0,x\lambda ;x,y)\\&
+(8 - 8\,y)\,G(\lambda ,0;0,y)\,G(\lambda ,x_0,x\lambda ;x,y)
+(8 - 8\,y)\,G(\lambda ,0;0,y)\,G(\lambda ,x_1,x\lambda ;x,y)\\&
+(8 - 8\,y)\,G(\lambda ,0;0,y)\,G(\lambda ,x\lambda ,0;x,y)
+8\,\left( -1 + y \right) \,G(\lambda ,0;0,y)\,G(x_0,\lambda ,x\lambda ;x,y)\\&
+8\,\left( -1 + y \right) \,G(\lambda ,0;0,y)\,G(x_0,x\lambda ,\lambda ;x,y)
+8\,\left( -1 + y \right) \,G(\lambda ,0;0,y)\,G(x_1,\lambda ,x\lambda ;x,y)\\&
+8\,\left( -1 + y \right) \,G(\lambda ,0;0,y)\,G(x_1,x\lambda ,\lambda ;x,y)
-16\,\left( -1 + y \right) \,G(\lambda ,0;0,y)\,G(x\lambda ,0,\lambda ;x,y)\\&
+(8 - 8\,y)\,G(\lambda ,0;0,y)\,G(x\lambda ,\lambda ,0;x,y)
+(8 - 8\,y)\,G(\lambda ,0;0,y)\,G(x\lambda ,x_0,\lambda ;x,y)\\&
+(8 - 8\,y)\,G(\lambda ,0;0,y)\,G(x\lambda ,x_1,\lambda ;x,y)
-16\,G(\lambda ,\lambda ;x,y)\,G(\lambda ,0,0;0,x_0)\\&
-16\,G(\lambda ,\lambda ;x,y)\,G(\lambda ,0,\lambda ;0,y)
-12\,G(\lambda ,\lambda ;x,y)\,G(\lambda ,\lambda ,0;0,x_0)\\&
+16\,G(\lambda ,\lambda ;x,y)\,G(\lambda ,\lambda ,0;0,y)
-8\,G(\lambda ,\lambda ;x,y)\,G(x_0,\lambda ,0;0,x_0)\\&
-8\,G(\lambda ,\lambda ;x,y)\,G(x_1,\lambda ,0;0,x_0)
+\frac{4\,{\pi }^2}{3}\,G(\lambda ,\lambda ;x,y)\,H(0;x_0)\\&
+\frac{-8\,{\pi }^2}{3}\,G(\lambda ,\lambda ;x,y)\,H(0;y)
+\frac{-4\,{\pi }^2}{3}\,G(\lambda ,\lambda ;x,y)\,H(1;x_0)\\&
+48\,G(\lambda ,\lambda ;x,y)\,H(0,0,0;x_0)
+12\,G(\lambda ,\lambda ;x,y)\,H(0,0,0;y)\\&
+8\,G(\lambda ,\lambda ;x,y)\,H(0,1,0;x_0)
-16\,G(\lambda ,\lambda ;x,y)\,H(0,1,0;y)\\&
+8\,G(\lambda ,\lambda ;x,y)\,H(1,0,0;x_0)
+16\,G(\lambda ,\lambda ;x,y)\,H(1,0,0;y)\\&
-8\,G(\lambda ,\lambda ;x,y)\,H(1,1,0;x_0)
-16\,\left( -1 + y \right) \,G(\lambda ,x\lambda ;x,y)\,G(\lambda ,0,0;0,y)\\&
+8\,{\left( -1 + y \right) }^2\,G(\lambda ,x\lambda ;x,y)\,G(\lambda ,0,x\lambda ;0,y)
+8\,{\left( -1 + y \right) }^2\,G(\lambda ,x\lambda ;x,y)\,G(\lambda ,x\lambda ,0;0,y)\\&
-16\,\left( -1 + y \right) \,G(\lambda ,x\lambda ;x,y)\,G(x_0,\lambda ,0;0,y)
-16\,\left( -1 + y \right) \,G(\lambda ,x\lambda ;x,y)\,G(x_1,\lambda ,0;0,y)\\&
+\frac{4\,{\pi }^2\,\left( -1 + y \right) }{3}\,G(\lambda ,x\lambda ;x,y)\,H(0;y)
+\frac{-8\,{\pi }^2\,\left( -1 + y \right) }{3}\,G(\lambda ,x\lambda ;x,y)\,H(1;y)\\&
+12\,\left( -1 + y \right) \,G(\lambda ,x\lambda ;x,y)\,H(0,0,0;y)
+8\,\left( -1 + y \right) \,G(\lambda ,x\lambda ;x,y)\,H(0,1,0;y)\\&
+24\,\left( -1 + y \right) \,G(\lambda ,x\lambda ;x,y)\,H(1,0,0;y)
-16\,\left( -1 + y \right) \,G(\lambda ,x\lambda ;x,y)\,H(1,1,0;y)\\&
-16\,\left( -1 + y \right) \,G(x\lambda ,\lambda ;x,y)\,G(\lambda ,0,0;0,y)
+8\,{\left( -1 + y \right) }^2\,G(x\lambda ,\lambda ;x,y)\,G(\lambda ,0,x\lambda ;0,y)\\&
+8\,{\left( -1 + y \right) }^2\,G(x\lambda ,\lambda ;x,y)\,G(\lambda ,x\lambda ,0;0,y)
-16\,\left( -1 + y \right) \,G(x\lambda ,\lambda ;x,y)\,G(x_0,\lambda ,0;0,y)\\&
-16\,\left( -1 + y \right) \,G(x\lambda ,\lambda ;x,y)\,G(x_1,\lambda ,0;0,y)
+\frac{4\,{\pi }^2\,\left( -1 + y \right) }{3}\,G(x\lambda ,\lambda ;x,y)\,H(0;y)\\&
+\frac{-8\,{\pi }^2\,\left( -1 + y \right) }{3}\,G(x\lambda ,\lambda ;x,y)\,H(1;y)
+12\,\left( -1 + y \right) \,G(x\lambda ,\lambda ;x,y)\,H(0,0,0;y)\\&
+8\,\left( -1 + y \right) \,G(x\lambda ,\lambda ;x,y)\,H(0,1,0;y)
+24\,\left( -1 + y \right) \,G(x\lambda ,\lambda ;x,y)\,H(1,0,0;y)\\&
-16\,\left( -1 + y \right) \,G(x\lambda ,\lambda ;x,y)\,H(1,1,0;y)
+16\,{\left( -1 + y \right) }^2\,G(x\lambda ,x\lambda ;x,y)\,G(\lambda ,0,0;0,x_0)\\&
+12\,{\left( -1 + y \right) }^2\,G(x\lambda ,x\lambda ;x,y)\,G(\lambda ,\lambda ,0;0,x_0)
+8\,{\left( -1 + y \right) }^2\,G(x\lambda ,x\lambda ;x,y)\,G(x_0,\lambda ,0;0,x_0)\\&
+8\,{\left( -1 + y \right) }^2\,G(x\lambda ,x\lambda ;x,y)\,G(x_1,\lambda ,0;0,x_0)
+\frac{-4\,{\pi }^2\,{\left( -1 + y \right) }^2}{3}\,G(x\lambda ,x\lambda ;x,y)\,H(0;x_0)\\&
+\frac{4\,{\pi }^2\,{\left( -1 + y \right) }^2}{3}\,G(x\lambda ,x\lambda ;x,y)\,H(1;x_0)
+12\,{\left( -1 + y \right) }^2\,G(x\lambda ,x\lambda ;x,y)\,H(0,0,0;y)\\&
-8\,{\left( -1 + y \right) }^2\,G(x\lambda ,x\lambda ;x,y)\,H(0,1,0;x_0)
-8\,{\left( -1 + y \right) }^2\,G(x\lambda ,x\lambda ;x,y)\,H(1,0,0;x_0)\\&
+8\,{\left( -1 + y \right) }^2\,G(x\lambda ,x\lambda ;x,y)\,H(1,1,0;x_0)
+24\,G(\lambda ,0,\lambda ;x,y)\,H(0,0;x_0)\\&
-4\,G(\lambda ,0,\lambda ;x,y)\,H(0,0;y)
+8\,G(\lambda ,0,\lambda ;x,y)\,H(1,0;x_0)\\&
+(4 - 4\,y)\,G(\lambda ,0,x\lambda ;x,y)\,H(0,0;y)
+8\,\left( -1 + y \right) \,G(\lambda ,0,x\lambda ;x,y)\,H(1,0;y)\\&
+48\,G(\lambda ,\lambda ,0;x,y)\,H(0,0;x_0)
-20\,G(\lambda ,\lambda ,0;x,y)\,H(0,0;y)\\&
+8\,G(\lambda ,\lambda ,0;x,y)\,H(1,0;x_0)
+16\,G(\lambda ,x_0,\lambda ;x,y)\,H(0,0;x_0)\\&
+4\,G(\lambda ,x_0,\lambda ;x,y)\,H(0,0;y)
+16\,G(\lambda ,x_0,\lambda ;x,y)\,H(1,0;x_0)\\&
+4\,\left( -1 + y \right) \,G(\lambda ,x_0,x\lambda ;x,y)\,H(0,0;y)
+8\,\left( -1 + y \right) \,G(\lambda ,x_0,x\lambda ;x,y)\,H(1,0;y)\\&
+16\,G(\lambda ,x_1,\lambda ;x,y)\,H(0,0;x_0)
+4\,G(\lambda ,x_1,\lambda ;x,y)\,H(0,0;y)\\&
+16\,G(\lambda ,x_1,\lambda ;x,y)\,H(1,0;x_0)
+4\,\left( -1 + y \right) \,G(\lambda ,x_1,x\lambda ;x,y)\,H(0,0;y)\\&
+8\,\left( -1 + y \right) \,G(\lambda ,x_1,x\lambda ;x,y)\,H(1,0;y)
+(8 - 8\,y)\,G(\lambda ,x\lambda ,0;x,y)\,H(0,0;y)\\&
+8\,\left( -1 + y \right) \,G(\lambda ,x\lambda ,0;x,y)\,H(1,0;y)
-16\,G(x_0,\lambda ,\lambda ;x,y)\,H(0,0;x_0)\\&
-4\,G(x_0,\lambda ,\lambda ;x,y)\,H(0,0;y)
-16\,G(x_0,\lambda ,\lambda ;x,y)\,H(1,0;x_0)\\&
+(4 - 4\,y)\,G(x_0,\lambda ,x\lambda ;x,y)\,H(0,0;y)
+(8 - 8\,y)\,G(x_0,\lambda ,x\lambda ;x,y)\,H(1,0;y)\\&
+(4 - 4\,y)\,G(x_0,x\lambda ,\lambda ;x,y)\,H(0,0;y)
+(8 - 8\,y)\,G(x_0,x\lambda ,\lambda ;x,y)\,H(1,0;y)\\&
-4\,{\left( -1 + y \right) }^2\,G(x_0,x\lambda ,x\lambda ;x,y)\,H(0,0;y)
-16\,G(x_1,\lambda ,\lambda ;x,y)\,H(0,0;x_0)\\&
-4\,G(x_1,\lambda ,\lambda ;x,y)\,H(0,0;y)
-16\,G(x_1,\lambda ,\lambda ;x,y)\,H(1,0;x_0)\\&
+(4 - 4\,y)\,G(x_1,\lambda ,x\lambda ;x,y)\,H(0,0;y)
+(8 - 8\,y)\,G(x_1,\lambda ,x\lambda ;x,y)\,H(1,0;y)\\&
+(4 - 4\,y)\,G(x_1,x\lambda ,\lambda ;x,y)\,H(0,0;y)
+(8 - 8\,y)\,G(x_1,x\lambda ,\lambda ;x,y)\,H(1,0;y)\\&
-4\,{\left( -1 + y \right) }^2\,G(x_1,x\lambda ,x\lambda ;x,y)\,H(0,0;y)
+8\,\left( -1 + y \right) \,G(x\lambda ,0,\lambda ;x,y)\,H(0,0;y)\\&
+16\,\left( -1 + y \right) \,G(x\lambda ,0,\lambda ;x,y)\,H(1,0;y)
+8\,{\left( -1 + y \right) }^2\,G(x\lambda ,0,x\lambda ;x,y)\,H(0,0;y)\\&
+(8 - 8\,y)\,G(x\lambda ,\lambda ,0;x,y)\,H(0,0;y)
+8\,\left( -1 + y \right) \,G(x\lambda ,\lambda ,0;x,y)\,H(1,0;y)\\&
+4\,\left( -1 + y \right) \,G(x\lambda ,x_0,\lambda ;x,y)\,H(0,0;y)
+8\,\left( -1 + y \right) \,G(x\lambda ,x_0,\lambda ;x,y)\,H(1,0;y)\\&
+4\,{\left( -1 + y \right) }^2\,G(x\lambda ,x_0,x\lambda ;x,y)\,H(0,0;y)
+4\,\left( -1 + y \right) \,G(x\lambda ,x_1,\lambda ;x,y)\,H(0,0;y)\\&
+8\,\left( -1 + y \right) \,G(x\lambda ,x_1,\lambda ;x,y)\,H(1,0;y)
+4\,{\left( -1 + y \right) }^2\,G(x\lambda ,x_1,x\lambda ;x,y)\,H(0,0;y)\\&
+4\,{\left( -1 + y \right) }^2\,G(x\lambda ,x\lambda ,0;x,y)\,H(0,0;y)
-8\,{\left( -1 + y \right) }^2\,G(x\lambda ,x\lambda ,0;x,y)\,H(1,0;x_0)\\&
+16\,G(\lambda ,0,0,\lambda ;x,y)\,H(0;x_0)
-8\,G(\lambda ,0,0,\lambda ;x,y)\,H(0;y)\\&
+(8 - 8\,y)\,G(\lambda ,0,0,x\lambda ;x,y)\,H(0;y)
+24\,G(\lambda ,0,\lambda ,0;x,y)\,H(0;x_0)\\&
-4\,G(\lambda ,0,\lambda ,0;x,y)\,H(0;y)
+8\,G(\lambda ,0,x_0,\lambda ;x,y)\,H(0;x_0)\\&
-4\,G(\lambda ,0,x_0,\lambda ;x,y)\,H(0;y)
+(4 - 4\,y)\,G(\lambda ,0,x_0,x\lambda ;x,y)\,H(0;y)\\&
+8\,G(\lambda ,0,x_1,\lambda ;x,y)\,H(0;x_0)
-4\,G(\lambda ,0,x_1,\lambda ;x,y)\,H(0;y)\\&
+(4 - 4\,y)\,G(\lambda ,0,x_1,x\lambda ;x,y)\,H(0;y)
+(4 - 4\,y)\,G(\lambda ,0,x\lambda ,0;x,y)\,H(0;y)\\&
+48\,G(\lambda ,\lambda ,0,0;x,y)\,H(0;x_0)
-8\,G(\lambda ,\lambda ,0,0;x,y)\,H(0;y)\\&
-72\,G(\lambda ,\lambda ,\lambda ,\lambda ;x,y)\,H(0;x_0)
+36\,G(\lambda ,\lambda ,\lambda ,\lambda ;x,y)\,H(0;y)\\&
+36\,\left( -1 + y \right) \,G(\lambda ,\lambda ,\lambda ,x\lambda ;x,y)\,H(0;y)
+8\,G(\lambda ,x_0,0,\lambda ;x,y)\,H(0;x_0)\\&
-4\,G(\lambda ,x_0,0,\lambda ;x,y)\,H(0;y)
+(4 - 4\,y)\,G(\lambda ,x_0,0,x\lambda ;x,y)\,H(0;y)\\&
+16\,G(\lambda ,x_0,\lambda ,0;x,y)\,H(0;x_0)
-8\,G(\lambda ,x_0,\lambda ,0;x,y)\,H(0;y)\\&
+(4 - 4\,y)\,G(\lambda ,x_0,x\lambda ,0;x,y)\,H(0;y)
+8\,G(\lambda ,x_1,0,\lambda ;x,y)\,H(0;x_0)\\&
-4\,G(\lambda ,x_1,0,\lambda ;x,y)\,H(0;y)
+(4 - 4\,y)\,G(\lambda ,x_1,0,x\lambda ;x,y)\,H(0;y)\\&
+16\,G(\lambda ,x_1,\lambda ,0;x,y)\,H(0;x_0)
-8\,G(\lambda ,x_1,\lambda ,0;x,y)\,H(0;y)\\&
+(4 - 4\,y)\,G(\lambda ,x_1,x\lambda ,0;x,y)\,H(0;y)
+24\,{\left( -1 + y \right) }^2\,G(\lambda ,x\lambda ,x\lambda ,\lambda ;x,y)\,H(0;x_0)\\&
-12\,{\left( -1 + y \right) }^2\,G(\lambda ,x\lambda ,x\lambda ,\lambda ;x,y)\,H(0;y)
-12\,{\left( -1 + y \right) }^3\,G(\lambda ,x\lambda ,x\lambda ,x\lambda ;x,y)\,H(0;y)\\&
-8\,G(x_0,\lambda ,0,\lambda ;x,y)\,H(0;x_0)
+4\,G(x_0,\lambda ,0,\lambda ;x,y)\,H(0;y)\\&
+4\,\left( -1 + y \right) \,G(x_0,\lambda ,0,x\lambda ;x,y)\,H(0;y)
-16\,G(x_0,\lambda ,\lambda ,0;x,y)\,H(0;x_0)\\&
+8\,G(x_0,\lambda ,\lambda ,0;x,y)\,H(0;y)
+4\,\left( -1 + y \right) \,G(x_0,\lambda ,x\lambda ,0;x,y)\,H(0;y)\\&
+4\,\left( -1 + y \right) \,G(x_0,x\lambda ,\lambda ,0;x,y)\,H(0;y)
-8\,G(x_1,\lambda ,0,\lambda ;x,y)\,H(0;x_0)\\&
+4\,G(x_1,\lambda ,0,\lambda ;x,y)\,H(0;y)
+4\,\left( -1 + y \right) \,G(x_1,\lambda ,0,x\lambda ;x,y)\,H(0;y)\\&
-16\,G(x_1,\lambda ,\lambda ,0;x,y)\,H(0;x_0)
+8\,G(x_1,\lambda ,\lambda ,0;x,y)\,H(0;y)\\&
+4\,\left( -1 + y \right) \,G(x_1,\lambda ,x\lambda ,0;x,y)\,H(0;y)
+4\,\left( -1 + y \right) \,G(x_1,x\lambda ,\lambda ,0;x,y)\,H(0;y)\\&
+(8 - 8\,y)\,G(x\lambda ,0,\lambda ,0;x,y)\,H(0;y)
+24\,{\left( -1 + y \right) }^2\,G(x\lambda ,\lambda ,x\lambda ,\lambda ;x,y)\,H(0;x_0)\\&
-12\,{\left( -1 + y \right) }^2\,G(x\lambda ,\lambda ,x\lambda ,\lambda ;x,y)\,H(0;y)
-12\,{\left( -1 + y \right) }^3\,G(x\lambda ,\lambda ,x\lambda ,x\lambda ;x,y)\,H(0;y)\\&
+(4 - 4\,y)\,G(x\lambda ,x_0,\lambda ,0;x,y)\,H(0;y)
+(4 - 4\,y)\,G(x\lambda ,x_1,\lambda ,0;x,y)\,H(0;y)\\&
+24\,{\left( -1 + y \right) }^2\,G(x\lambda ,x\lambda ,\lambda ,\lambda ;x,y)\,H(0;x_0)
-12\,{\left( -1 + y \right) }^2\,G(x\lambda ,x\lambda ,\lambda ,\lambda ;x,y)\,H(0;y)\\&
-12\,{\left( -1 + y \right) }^3\,G(x\lambda ,x\lambda ,\lambda ,x\lambda ;x,y)\,H(0;y)
+4\,\Delta G(0;0,x_0)\,G(\lambda ,0;0,x_0)\,G(\lambda ,\lambda ;x,y)\\&
-4\,{\left( -1 + y \right) }^2\,\Delta G(0;0,x_0)\,G(\lambda ,0;0,x_0)\,G(x\lambda ,x\lambda ;x,y)
+4\,\Delta G(0;0,x_0)\,G(\lambda ,\lambda ;x,y)\,H(0,0;x_0)\\&
-4\,\Delta G(0;0,x_0)\,G(\lambda ,\lambda ;x,y)\,H(1,0;x_0)
-4\,{\left( -1 + y \right) }^2\,\Delta G(0;0,x_0)\,G(x\lambda ,x\lambda ;x,y)\,H(0,0;x_0)\\&
+4\,{\left( -1 + y \right) }^2\,\Delta G(0;0,x_0)\,G(x\lambda ,x\lambda ;x,y)\,H(1,0;x_0)
+4\,\Delta G(0;0,x_0)\,G(\lambda ,0,\lambda ;x,y)\,H(0;x_0)\\&
+4\,\Delta G(0;0,x_0)\,G(\lambda ,\lambda ,0;x,y)\,H(0;x_0)
+8\,\Delta G(0;0,x_0)\,G(\lambda ,x_0,\lambda ;x,y)\,H(0;x_0)\\&
+8\,\Delta G(0;0,x_0)\,G(\lambda ,x_1,\lambda ;x,y)\,H(0;x_0)
-8\,\Delta G(0;0,x_0)\,G(x_0,\lambda ,\lambda ;x,y)\,H(0;x_0)\\&
-8\,\Delta G(0;0,x_0)\,G(x_1,\lambda ,\lambda ;x,y)\,H(0;x_0)
-4\,{\left( -1 + y \right) }^2\,\Delta G(0;0,x_0)\,G(x\lambda ,x\lambda ,0;x,y)\,H(0;x_0)\\&
+8\,\Delta G(0;0,y)\,G(\lambda ,\lambda ;x,y)\,H(0,0;y)
+12\,\left( -1 + y \right) \,\Delta G(0;0,y)\,G(\lambda ,x\lambda ;x,y)\,H(0,0;y)\\&
+(8 - 8\,y)\,\Delta G(0;0,y)\,G(\lambda ,x\lambda ;x,y)\,H(1,0;y)
+12\,\left( -1 + y \right) \,\Delta G(0;0,y)\,G(x\lambda ,\lambda ;x,y)\,H(0,0;y)\\&
+(8 - 8\,y)\,\Delta G(0;0,y)\,G(x\lambda ,\lambda ;x,y)\,H(1,0;y)
+4\,\left( -1 + y \right) \,\Delta G(0;0,y)\,G(\lambda ,0,x\lambda ;x,y)\,H(0;y)\\&
+4\,\left( -1 + y \right) \,\Delta G(0;0,y)\,G(\lambda ,x_0,x\lambda ;x,y)\,H(0;y)
+4\,\left( -1 + y \right) \,\Delta G(0;0,y)\,G(\lambda ,x_1,x\lambda ;x,y)\,H(0;y)\\&
+4\,\left( -1 + y \right) \,\Delta G(0;0,y)\,G(\lambda ,x\lambda ,0;x,y)\,H(0;y)
+(4 - 4\,y)\,\Delta G(0;0,y)\,G(x_0,\lambda ,x\lambda ;x,y)\,H(0;y)\\&
+(4 - 4\,y)\,\Delta G(0;0,y)\,G(x_0,x\lambda ,\lambda ;x,y)\,H(0;y)
+(4 - 4\,y)\,\Delta G(0;0,y)\,G(x_1,\lambda ,x\lambda ;x,y)\,H(0;y)\\&
+(4 - 4\,y)\,\Delta G(0;0,y)\,G(x_1,x\lambda ,\lambda ;x,y)\,H(0;y)
+8\,\left( -1 + y \right) \,\Delta G(0;0,y)\,G(x\lambda ,0,\lambda ;x,y)\,H(0;y)\\&
+4\,\left( -1 + y \right) \,\Delta G(0;0,y)\,G(x\lambda ,\lambda ,0;x,y)\,H(0;y)
+4\,\left( -1 + y \right) \,\Delta G(0;0,y)\,G(x\lambda ,x_0,\lambda ;x,y)\,H(0;y)\\&
+4\,\left( -1 + y \right) \,\Delta G(0;0,y)\,G(x\lambda ,x_1,\lambda ;x,y)\,H(0;y)
+2\,\Delta G(0,0;0,x_0)\,G(\lambda ,\lambda ;x,y)\,H(0;x_0)\\&
-2\,{\left( -1 + y \right) }^2\,\Delta G(0,0;0,x_0)\,G(x\lambda ,x\lambda ;x,y)\,H(0;x_0)
+(4 - 4\,y)\,\Delta G(0,0;0,y)\,G(\lambda ,x\lambda ;x,y)\,H(0;y)\\&
+(4 - 4\,y)\,\Delta G(0,0;0,y)\,G(x\lambda ,\lambda ;x,y)\,H(0;y)
+6\,\Delta G(0,\lambda ;0,x_0)\,G(\lambda ,\lambda ;x,y)\,H(0;x_0)\\&
-12\,\Delta G(0,\lambda ;0,x_0)\,G(\lambda ,\lambda ;x,y)\,H(0;y)
-6\,{\left( -1 + y \right) }^2\,\Delta G(0,\lambda ;0,x_0)\,G(x\lambda ,x\lambda ;x,y)\,H(0;x_0)\\&
+12\,{\left( -1 + y \right) }^2\,\Delta G(0,\lambda ;0,x_0)\,G(x\lambda ,x\lambda ;x,y)\,H(0;y)
+8\,\Delta G(0,\lambda ;0,y)\,G(\lambda ,\lambda ;x,y)\,H(0;y)\\&
-16\,\left( -1 + y \right) \,\Delta G(0,\lambda ;0,y)\,G(\lambda ,x\lambda ;x,y)\,H(0;x_0)
+8\,\left( -1 + y \right) \,\Delta G(0,\lambda ;0,y)\,G(\lambda ,x\lambda ;x,y)\,H(0;y)\\&
-16\,\left( -1 + y \right) \,\Delta G(0,\lambda ;0,y)\,G(x\lambda ,\lambda ;x,y)\,H(0;x_0)
+8\,\left( -1 + y \right) \,\Delta G(0,\lambda ;0,y)\,G(x\lambda ,\lambda ;x,y)\,H(0;y)\\&
+6\,\left( -2\,{\sqrt{y}} + y \right) \,\Delta G(0,x\lambda ;0,x_0)\,G(\lambda ,\lambda ;x,y)\,H(0;x_0)\\&
-6\,\left( -2 + {\sqrt{y}} \right) \,{\left( -1 + y \right) }^2\,{\sqrt{y}}\,\Delta G(0,x\lambda ;0,x_0)\,G(x\lambda ,x\lambda ;x,y)\,H(0;x_0)\\&
+4\,{\left( -1 + y \right) }^2\,\Delta G(0,x\lambda ;0,y)\,G(\lambda ,x\lambda ;x,y)\,H(0;y)
+4\,{\left( -1 + y \right) }^2\,\Delta G(0,x\lambda ;0,y)\,G(x\lambda ,\lambda ;x,y)\,H(0;y)\\&
+4\,G(\lambda ;0,x_0)\,G(\lambda ,0;0,x_0)\,G(\lambda ,\lambda ;x,y)
-4\,{\left( -1 + y \right) }^2\,G(\lambda ;0,x_0)\,G(\lambda ,0;0,x_0)\,G(x\lambda ,x\lambda ;x,y)\\&
+4\,G(\lambda ;0,x_0)\,G(\lambda ,\lambda ;x,y)\,H(0,0;x_0)
-16\,G(\lambda ;0,x_0)\,G(\lambda ,\lambda ;x,y)\,H(0,0;y)\\&
-4\,G(\lambda ;0,x_0)\,G(\lambda ,\lambda ;x,y)\,H(1,0;x_0)
-4\,{\left( -1 + y \right) }^2\,G(\lambda ;0,x_0)\,G(x\lambda ,x\lambda ;x,y)\,H(0,0;x_0)\\&
+16\,{\left( -1 + y \right) }^2\,G(\lambda ;0,x_0)\,G(x\lambda ,x\lambda ;x,y)\,H(0,0;y)
+4\,{\left( -1 + y \right) }^2\,G(\lambda ;0,x_0)\,G(x\lambda ,x\lambda ;x,y)\,H(1,0;x_0)\\&
+4\,G(\lambda ;0,x_0)\,G(\lambda ,0,\lambda ;x,y)\,H(0;x_0)
-8\,G(\lambda ;0,x_0)\,G(\lambda ,0,\lambda ;x,y)\,H(0;y)\\&
+4\,G(\lambda ;0,x_0)\,G(\lambda ,\lambda ,0;x,y)\,H(0;x_0)
-8\,G(\lambda ;0,x_0)\,G(\lambda ,\lambda ,0;x,y)\,H(0;y)\\&
+8\,G(\lambda ;0,x_0)\,G(\lambda ,x_0,\lambda ;x,y)\,H(0;x_0)
-16\,G(\lambda ;0,x_0)\,G(\lambda ,x_0,\lambda ;x,y)\,H(0;y)\\&
+8\,G(\lambda ;0,x_0)\,G(\lambda ,x_1,\lambda ;x,y)\,H(0;x_0)
-16\,G(\lambda ;0,x_0)\,G(\lambda ,x_1,\lambda ;x,y)\,H(0;y)\\&
-8\,G(\lambda ;0,x_0)\,G(x_0,\lambda ,\lambda ;x,y)\,H(0;x_0)
+16\,G(\lambda ;0,x_0)\,G(x_0,\lambda ,\lambda ;x,y)\,H(0;y)\\&
-8\,G(\lambda ;0,x_0)\,G(x_1,\lambda ,\lambda ;x,y)\,H(0;x_0)
+16\,G(\lambda ;0,x_0)\,G(x_1,\lambda ,\lambda ;x,y)\,H(0;y)\\&
-4\,{\left( -1 + y \right) }^2\,G(\lambda ;0,x_0)\,G(x\lambda ,x\lambda ,0;x,y)\,H(0;x_0)
+8\,{\left( -1 + y \right) }^2\,G(\lambda ;0,x_0)\,G(x\lambda ,x\lambda ,0;x,y)\,H(0;y)\\&
+8\,G(\lambda ;0,y)\,G(\lambda ,\lambda ;x,y)\,H(0,0;y)
+16\,G(\lambda ;0,y)\,G(\lambda ,\lambda ;x,y)\,H(1,0;y)\\&
-32\,\left( -1 + y \right) \,G(\lambda ;0,y)\,G(\lambda ,x\lambda ;x,y)\,H(0,0;x_0)
+12\,\left( -1 + y \right) \,G(\lambda ;0,y)\,G(\lambda ,x\lambda ;x,y)\,H(0,0;y)\\&
-32\,\left( -1 + y \right) \,G(\lambda ;0,y)\,G(x\lambda ,\lambda ;x,y)\,H(0,0;x_0)
+12\,\left( -1 + y \right) \,G(\lambda ;0,y)\,G(x\lambda ,\lambda ;x,y)\,H(0,0;y)\\&
+(8 - 8\,y)\,G(\lambda ;0,y)\,G(\lambda ,0,x\lambda ;x,y)\,H(0;x_0)
+4\,\left( -1 + y \right) \,G(\lambda ;0,y)\,G(\lambda ,0,x\lambda ;x,y)\,H(0;y)\\&
+(8 - 8\,y)\,G(\lambda ;0,y)\,G(\lambda ,x_0,x\lambda ;x,y)\,H(0;x_0)
+4\,\left( -1 + y \right) \,G(\lambda ;0,y)\,G(\lambda ,x_0,x\lambda ;x,y)\,H(0;y)\\&
+(8 - 8\,y)\,G(\lambda ;0,y)\,G(\lambda ,x_1,x\lambda ;x,y)\,H(0;x_0)
+4\,\left( -1 + y \right) \,G(\lambda ;0,y)\,G(\lambda ,x_1,x\lambda ;x,y)\,H(0;y)\\&
+(8 - 8\,y)\,G(\lambda ;0,y)\,G(\lambda ,x\lambda ,0;x,y)\,H(0;x_0)
+4\,\left( -1 + y \right) \,G(\lambda ;0,y)\,G(\lambda ,x\lambda ,0;x,y)\,H(0;y)\\&
+8\,\left( -1 + y \right) \,G(\lambda ;0,y)\,G(x_0,\lambda ,x\lambda ;x,y)\,H(0;x_0)
+(4 - 4\,y)\,G(\lambda ;0,y)\,G(x_0,\lambda ,x\lambda ;x,y)\,H(0;y)\\&
+8\,\left( -1 + y \right) \,G(\lambda ;0,y)\,G(x_0,x\lambda ,\lambda ;x,y)\,H(0;x_0)
+(4 - 4\,y)\,G(\lambda ;0,y)\,G(x_0,x\lambda ,\lambda ;x,y)\,H(0;y)\\&
+8\,\left( -1 + y \right) \,G(\lambda ;0,y)\,G(x_1,\lambda ,x\lambda ;x,y)\,H(0;x_0)
+(4 - 4\,y)\,G(\lambda ;0,y)\,G(x_1,\lambda ,x\lambda ;x,y)\,H(0;y)\\&
+8\,\left( -1 + y \right) \,G(\lambda ;0,y)\,G(x_1,x\lambda ,\lambda ;x,y)\,H(0;x_0)
+(4 - 4\,y)\,G(\lambda ;0,y)\,G(x_1,x\lambda ,\lambda ;x,y)\,H(0;y)\\&
-16\,\left( -1 + y \right) \,G(\lambda ;0,y)\,G(x\lambda ,0,\lambda ;x,y)\,H(0;x_0)
+8\,\left( -1 + y \right) \,G(\lambda ;0,y)\,G(x\lambda ,0,\lambda ;x,y)\,H(0;y)\\&
+(8 - 8\,y)\,G(\lambda ;0,y)\,G(x\lambda ,\lambda ,0;x,y)\,H(0;x_0)
+4\,\left( -1 + y \right) \,G(\lambda ;0,y)\,G(x\lambda ,\lambda ,0;x,y)\,H(0;y)\\&
+(8 - 8\,y)\,G(\lambda ;0,y)\,G(x\lambda ,x_0,\lambda ;x,y)\,H(0;x_0)
+4\,\left( -1 + y \right) \,G(\lambda ;0,y)\,G(x\lambda ,x_0,\lambda ;x,y)\,H(0;y)\\&
+(8 - 8\,y)\,G(\lambda ;0,y)\,G(x\lambda ,x_1,\lambda ;x,y)\,H(0;x_0)
+4\,\left( -1 + y \right) \,G(\lambda ;0,y)\,G(x\lambda ,x_1,\lambda ;x,y)\,H(0;y)\\&
+2\,G(\lambda ,0;0,x_0)\,G(\lambda ,\lambda ;x,y)\,H(0;x_0)
-16\,G(\lambda ,0;0,x_0)\,G(\lambda ,\lambda ;x,y)\,H(0;y)\\&
-2\,{\left( -1 + y \right) }^2\,G(\lambda ,0;0,x_0)\,G(x\lambda ,x\lambda ;x,y)\,H(0;x_0)
+16\,{\left( -1 + y \right) }^2\,G(\lambda ,0;0,x_0)\,G(x\lambda ,x\lambda ;x,y)\,H(0;y)\\&
-24\,\left( -1 + y \right) \,G(\lambda ,0;0,y)\,G(\lambda ,x\lambda ;x,y)\,H(0;x_0)
+(4 - 4\,y)\,G(\lambda ,0;0,y)\,G(\lambda ,x\lambda ;x,y)\,H(0;y)\\&
-24\,\left( -1 + y \right) \,G(\lambda ,0;0,y)\,G(x\lambda ,\lambda ;x,y)\,H(0;x_0)
+(4 - 4\,y)\,G(\lambda ,0;0,y)\,G(x\lambda ,\lambda ;x,y)\,H(0;y)\\&
+6\,G(\lambda ,\lambda ;0,x_0)\,G(\lambda ,\lambda ;x,y)\,H(0;x_0)
-12\,G(\lambda ,\lambda ;0,x_0)\,G(\lambda ,\lambda ;x,y)\,H(0;y)\\&
-6\,{\left( -1 + y \right) }^2\,G(\lambda ,\lambda ;0,x_0)\,G(x\lambda ,x\lambda ;x,y)\,H(0;x_0)
+12\,{\left( -1 + y \right) }^2\,G(\lambda ,\lambda ;0,x_0)\,G(x\lambda ,x\lambda ;x,y)\,H(0;y)\\&
+16\,G(\lambda ,\lambda ;0,y)\,G(\lambda ,\lambda ;x,y)\,H(0;x_0)
-8\,G(\lambda ,\lambda ;0,y)\,G(\lambda ,\lambda ;x,y)\,H(0;y)\\&
+6\,\left( -2\,{\sqrt{y}} + y \right) \,G(\lambda ,\lambda ;x,y)\,G(\lambda ,x\lambda ;0,x_0)\,H(0;x_0)\\&
-16\,\left( -1 + y \right) \,G(\lambda ,\lambda ;x,y)\,G(\lambda ,x\lambda ;0,y)\,H(0;y)\\&
+4\,G(\lambda ,\lambda ;x,y)\,G(x_0,\lambda ;0,x_0)\,H(0;x_0)
-8\,G(\lambda ,\lambda ;x,y)\,G(x_0,\lambda ;0,x_0)\,H(0;y)\\&
+4\,\left( -2\,{\sqrt{y}} + y \right) \,G(\lambda ,\lambda ;x,y)\,G(x_0,x\lambda ;0,x_0)\,H(0;x_0)\\&
+4\,G(\lambda ,\lambda ;x,y)\,G(x_1,\lambda ;0,x_0)\,H(0;x_0)
-8\,G(\lambda ,\lambda ;x,y)\,G(x_1,\lambda ;0,x_0)\,H(0;y)\\&
+4\,\left( -2\,{\sqrt{y}} + y \right) \,G(\lambda ,\lambda ;x,y)\,G(x_1,x\lambda ;0,x_0)\,H(0;x_0)\\&
-4\,G(\lambda ,\lambda ;x,y)\,H(0;x_0)\,H(0,0;y)
-4\,G(\lambda ,\lambda ;x,y)\,H(0;y)\,H(0,0;x_0)\\&
-6\,\left( -2 + {\sqrt{y}} \right) \,{\left( -1 + y \right) }^2\,{\sqrt{y}}\,G(\lambda ,x\lambda ;0,x_0)\,G(x\lambda ,x\lambda ;x,y)\,H(0;x_0)\\&
+8\,{\left( -1 + y \right) }^2\,G(\lambda ,x\lambda ;0,y)\,G(\lambda ,x\lambda ;x,y)\,H(0;x_0)
-4\,{\left( -1 + y \right) }^2\,G(\lambda ,x\lambda ;0,y)\,G(\lambda ,x\lambda ;x,y)\,H(0;y)\\&
+8\,{\left( -1 + y \right) }^2\,G(\lambda ,x\lambda ;0,y)\,G(x\lambda ,\lambda ;x,y)\,H(0;x_0)
-4\,{\left( -1 + y \right) }^2\,G(\lambda ,x\lambda ;0,y)\,G(x\lambda ,\lambda ;x,y)\,H(0;y)\\&
-16\,\left( -1 + y \right) \,G(\lambda ,x\lambda ;x,y)\,G(x_0,\lambda ;0,y)\,H(0;x_0)
+8\,\left( -1 + y \right) \,G(\lambda ,x\lambda ;x,y)\,G(x_0,\lambda ;0,y)\,H(0;y)\\&
+8\,{\left( -1 + y \right) }^2\,G(\lambda ,x\lambda ;x,y)\,G(x_0,x\lambda ;0,y)\,H(0;y)
-16\,\left( -1 + y \right) \,G(\lambda ,x\lambda ;x,y)\,G(x_1,\lambda ;0,y)\,H(0;x_0)\\&
+8\,\left( -1 + y \right) \,G(\lambda ,x\lambda ;x,y)\,G(x_1,\lambda ;0,y)\,H(0;y)
+8\,{\left( -1 + y \right) }^2\,G(\lambda ,x\lambda ;x,y)\,G(x_1,x\lambda ;0,y)\,H(0;y)\\&
+4\,\left( -1 + y \right) \,G(\lambda ,x\lambda ;x,y)\,H(0;x_0)\,H(0,0;y)
+4\,\left( -1 + y \right) \,G(\lambda ,x\lambda ;x,y)\,H(0;y)\,H(0,0;x_0)\\&
-4\,{\left( -1 + y \right) }^2\,G(x_0,\lambda ;0,x_0)\,G(x\lambda ,x\lambda ;x,y)\,H(0;x_0)
+8\,{\left( -1 + y \right) }^2\,G(x_0,\lambda ;0,x_0)\,G(x\lambda ,x\lambda ;x,y)\,H(0;y)\\&
-16\,\left( -1 + y \right) \,G(x_0,\lambda ;0,y)\,G(x\lambda ,\lambda ;x,y)\,H(0;x_0)
+8\,\left( -1 + y \right) \,G(x_0,\lambda ;0,y)\,G(x\lambda ,\lambda ;x,y)\,H(0;y)\\&
-4\,\left( -2 + {\sqrt{y}} \right) \,{\left( -1 + y \right) }^2\,{\sqrt{y}}\,G(x_0,x\lambda ;0,x_0)\,G(x\lambda ,x\lambda ;x,y)\,H(0;x_0)\\&
+8\,{\left( -1 + y \right) }^2\,G(x_0,x\lambda ;0,y)\,G(x\lambda ,\lambda ;x,y)\,H(0;y)
-4\,{\left( -1 + y \right) }^2\,G(x_1,\lambda ;0,x_0)\,G(x\lambda ,x\lambda ;x,y)\,H(0;x_0)\\&
+8\,{\left( -1 + y \right) }^2\,G(x_1,\lambda ;0,x_0)\,G(x\lambda ,x\lambda ;x,y)\,H(0;y)
-16\,\left( -1 + y \right) \,G(x_1,\lambda ;0,y)\,G(x\lambda ,\lambda ;x,y)\,H(0;x_0)\\&
+8\,\left( -1 + y \right) \,G(x_1,\lambda ;0,y)\,G(x\lambda ,\lambda ;x,y)\,H(0;y)\\&
-4\,\left( -2 + {\sqrt{y}} \right) \,{\left( -1 + y \right) }^2\,{\sqrt{y}}\,G(x_1,x\lambda ;0,x_0)\,G(x\lambda ,x\lambda ;x,y)\,H(0;x_0)\\&
+8\,{\left( -1 + y \right) }^2\,G(x_1,x\lambda ;0,y)\,G(x\lambda ,\lambda ;x,y)\,H(0;y)
+4\,\left( -1 + y \right) \,G(x\lambda ,\lambda ;x,y)\,H(0;x_0)\,H(0,0;y)\\&
+4\,\left( -1 + y \right) \,G(x\lambda ,\lambda ;x,y)\,H(0;y)\,H(0,0;x_0)
-4\,{\left( -1 + y \right) }^2\,G(x\lambda ,x\lambda ;x,y)\,H(0;x_0)\,H(0,0;y)\\&
-4\,{\left( -1 + y \right) }^2\,G(x\lambda ,x\lambda ;x,y)\,H(0;y)\,H(0,0;x_0)
-4\,G(\lambda ,\lambda ,0;x,y)\,H(0;x_0)\,H(0;y)\\&
+4\,\left( -1 + y \right) \,G(\lambda ,x\lambda ,0;x,y)\,H(0;x_0)\,H(0;y)
+4\,\left( -1 + y \right) \,G(x\lambda ,\lambda ,0;x,y)\,H(0;x_0)\,H(0;y)\\&
-4\,{\left( -1 + y \right) }^2\,G(x\lambda ,x\lambda ,0;x,y)\,H(0;x_0)\,H(0;y)

  .\\
\end{align*}
As with the other MI, we have checked that the leading
(finite) contribution agrees with 
the results of Ref.~\cite{finitetwoloop}.

\section{Summary}
\label{sec:summary}

In this paper, we have provided series expansions in the dimensional
regularisation parameter $\epsilon$ for all two-loop Master Integrals with
three external off-shell legs and all internal  lines being massless. The
results are presented in terms of an extended basis of 2-dimensional
harmonic polylogarithms. The novel feature is that this basis includes
quadratic forms - that matches on to the allowed  phase space boundary for
the $1\to 2$ decay. For each Master Integral, we have given sufficient
terms in the $\epsilon$-expansion to describe two-loop vertex corrections
for physical processes.

The MI presented here are ingredients for a variety of interesting
two-loop processes such as the QCD corrections to $H \to V^*V^*$ decay in
the heavy top quark limit  and the QCD corrections to the fully off-shell
triple gluon (and quark-gluon) vertices. 

The MI also form a staging post for the study of massless two-loop  $2\to
2$ scattering amplitudes with two off-shell legs. These processes include
the NNLO QCD corrections to $q\bar q \to V^* V^*$ (where $V = W$,$Z$) and
the NLO corrections to $gg \to V^*V^*$.   Altogether there are 11 planar
box and 3 non-planar box master topologies which remain to be studied.

\section{Acknowledgements}
\label{sec:ack}

This work was supported in part by the UK Particle Physics and Astronomy 
Research Council and by the EU Fifth Framework Programme `Improving Human
Potential', Research Training Network `Particle Physics Phenomenology  at
High Energy Colliders', contract HPRN-CT-2000-00149.

\end{fmffile}

\end{document}